\documentclass{article}
\usepackage[utf8]{inputenc}
\usepackage{fullpage}
\usepackage{xcolor}

\usepackage{graphicx}
\usepackage{amsmath}
\usepackage{amssymb}
\usepackage{mathrsfs}
\usepackage{url}
\usepackage{tikz}

\definecolor{myred}{HTML}{872A34}
\definecolor{mypink}{HTML}{E86875}

\newcommand*{\mysquare}[1]{\tikz{\filldraw[draw=#1,fill=#1] (0,0)
rectangle (0.2cm,0.2cm);}}

\title{The why, how, and when of representations for complex systems}

\author{
  Leo Torres \\
  \texttt{leo@leotrs.com} \\
  Network Science Institute, \\
  Northeastern University \\
  \and
  Ann S. Blevins \\
  \texttt{annsize@seas.upenn.edu} \\
  Department of Bioengineering, \\
  University of Pennsylvania \\
  \and
  Danielle S. Bassett \\
  \texttt{dsb@seas.upenn.edu} \\
  Department of Bioengineering, \\
  University of Pennsylvania \\
  \and
  Tina Eliassi-Rad \\
  \texttt{tina@eliassi.org} \\
  Network Science Institute and \\
  Khoury College of Computer Sciences, \\
  Northeastern University
}

\date{\today}

\begin{document}
\maketitle

\newpage
\tableofcontents

\newpage 
\begin{abstract}
    Complex systems thinking is applied to a wide variety of domains, from neuroscience to computer science and economics. The wide variety of implementations has resulted in two key challenges: the progenation of many domain-specific strategies that are seldom revisited or questioned, and the siloing of ideas within a domain due to inconsistency of complex systems language. In this work we offer basic, domain-agnostic language in order to advance towards a more cohesive vocabulary. We use this language to evaluate each step of the complex systems analysis pipeline, beginning with the system and data collected, then moving through different mathematical formalisms for encoding the observed data (i.e. graphs, simplicial complexes, and hypergraphs), and relevant computational methods for each formalism. At each step we consider different types of \emph{dependencies}; these are properties of the system that describe how the existence of one relation among the parts of a system may influence the existence of another relation. We discuss how dependencies may arise and how they may alter interpretation of results or the entirety of the analysis pipeline. We close with two real-world examples using coauthorship data and email communications data that illustrate how the system under study, the dependencies therein, the research question, and choice of mathematical representation influence the results. We hope this work can serve as an opportunity of reflection for experienced complexity scientists, as well as an introductory resource for new researchers.
\end{abstract}

\newpage
\section{Introduction}\label{sec:introduction}

The term ``complex system'' is used to describe a multitude of systems of markedly different magnitudes, from the atomic scale of interacting atoms to the vast scale of the whole universe, as well as markedly different behaviors, from starling murmurations to the viral spread of information on social media. Though distinct definitions exist, and not one is globally agreed upon, in general a complex system is a collection of objects or agents that (a) have a high cardinality and (b) interact with one another in a non-trivial way, such that (c) the collective behavior of the system is unexpected, different than, or not immediately predictable from the aggregation of the behavior of the individual parts. This unique collective behavior is often said to \emph{emerge} from the dynamics of the parts \cite{johnson2006emergent,kivelson2016defining}. Real world examples include computations in a neuronal population, cellular reactions in photosynthesis, food webs in ecology, transactions in local markets, interconnected world-wide trading in economics, and various technologies such as the Internet and the power grid.

In order to study complex systems across disciplines and domains, a first step is to concretely represent the system using a unifying mathematical language. In recent decades, the discipline of network science has arisen as the main focus of development of such a language \cite{newman2018networks}. Network scientists typically study complex systems by first modeling them using the tools and frameworks afforded by disciplines such as discrete mathematics and computational data structures. These formal frameworks enable the application of tried and true methodologies coming from different subfields within the mathematical, physical, and computational sciences. Furthermore these formalisms allow for the execution of efficient algorithms and can be used to infer structure, function, and dynamics of a system. What makes this process somewhat challenging is that each encounter with a new complex system requires the construction of a new representation tailored to it. Network science is far from developing a single, unified language that allows the study of all possible system structures and behaviors \cite{ladyman2020what}. Indeed, there is currently not one, but a wealth of related frameworks, each of which captures particular perspectives and properties of the system under study. 

This wealth of frameworks, and the resulting wealth of accompanying analysis pipelines, creates challenges for the study of complex systems. It hinders interdisciplinary communication, as researchers in one discipline may be unfamiliar with the frameworks and procedures used in another. Even within a single subfield, various approaches to represent and analyze the same complex system can hinder collective insight across research groups or projects. As a consequence, it is difficult and sometimes impossible to gather insight across systems, which directly hampers the progress of complexity science \cite{mitchell2009complexity}. As conscientious researchers, we must address this challenge by understanding the assumptions underlying each formalism, as well as the the relationships between formalisms, and the impact of both formalism assumptions and relations on our analyses and interpretations of results.

\begin{figure}
    \centering
    \includegraphics{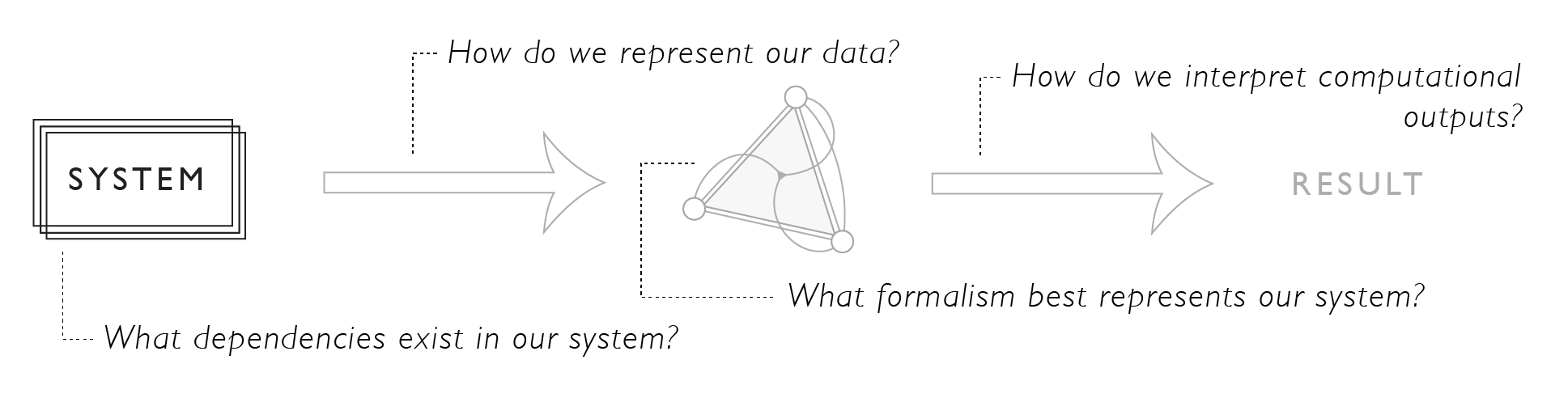}
    \caption{\textbf{Linear flow of a basic analysis pipeline for complex systems.} We begin with the system under study, and ask what sorts of elementary units exist, what relations exist that group elements together, and what dependencies might influence the existence of relations among units. We then turn to the question of how to represent the units, relations, and their dependencies; to answer this question, we must choose a formalism. Finally, we seek to interpret the outcomes of computations performed on the representation, and from those interpretations we reach a conclusion about the structure and function of the system.} 
    \label{fig:flow1}
\end{figure}

In this work we aim to collect and align complex system analysis pipelines while providing a common vocabulary for a continued discussion. While achieving a single, unified language is unlikely, we can at the very least begin to simplify and condense the frameworks currently in use. For clarity, we begin by defining the fundamental terms used throughout the paper. The main text follows the flow of Fig.~\ref{fig:flow1}, which illustrates a simplified representation of the analysis pipeline used when studying a complex system, insofar as it pertains to the formal representation of the system. We begin with an investigation of common system properties which we call \emph{dependencies}, followed by definitions of three mathematical formalisms commonly used for representation. Next we highlight mathematical relationships between formalisms that one might utilize in order to answer particular research questions, and finally we provide examples of computations suited for each of the three formalisms. Throughout the text we repeatedly ask how these dependencies and other modeling choices may influence the pipeline steps discussed. We provide two examples using a co-authorship dataset and the Enron emails dataset \cite{BensonASJK18} to demonstrate the effects of various analysis pipelines on the results obtained from the same underlying system. Finally we close by suggesting that each modeling decision in a research analysis pipeline be taken on a case-by-case basis and in consideration of the dependencies, formalisms, relationships, and research questions. We hope this paper can serve as an instructive resource for new researchers in the field, and an opportunity for reflection to those more experienced in complex systems analyses. 

\subsection{Definitions}\label{sec:definitions}

In this work we use a consistent language to allow for effective and precise communication between scientists across disciplines. Here we provide a list of terms that we will use throughout this paper and their definitions. With the condensing of vocabulary and precise definitions of often abstract concepts, we hope to operationalize the study of the structure and behavior of complex systems.

\begin{itemize}
\item \textbf{Unit, element, or node:} an individual object, agent, or part of a system. Unless otherwise specified, we denote the set of nodes by $V$.

\item \textbf{Relation:} a set $r$ of one or more nodes, such that $r\subseteq V$. In practice, node relations can arise from correlations in data, observed interactions between units, or groups of elements known to function collectively. A relation $r$ can be \emph{dyadic} if it contains exactly two units ($|r|=2)$, or \emph{polyadic} if the relation contains three or more units ($|r|>2$). If $r$ contains $k$ nodes, then we say those $k$ nodes in $r$ are related. In some parts of the literature, polyadic relations have also been called ``higher order'' relations, and have been used to refer to motifs in graphs \cite{benson2016higher}. To avoid confusion, however, in this paper we will use ``higher order'' to refer exclusively to a particular formalism introduced in Section \ref{sec:multiple_dep}. We denote the set of relations by $R$ unless a domain-specific convention already exists. 

\item \textbf{System:} a collection of units $V$ and all relations $R$, such that the collection needs no other pieces in order to function completely or to interact autonomously with its environment. The set of units are the components of the system, while the patterns found in the set of relations are called the system's \emph{structure}. The system's activity, including changes in nodes and relations over time, is sometimes also called its \emph{function}.

\item \textbf{Complex system:} a system whose units and relations together exhibit a qualitatively different functionality than the sum of its units acting individually; the subject of study of complexity science. In this work, ``system'' always refers to a complex system.

\item \textbf{System fragment:} a subset of the nodes and relations of a system. Researchers usually do not have access to all units or all relevant relations. Instead, they usually have access to --and must perform their studies on-- fragments of a system. Sometimes this limited access is due to the vast number of units (a human brain contains on the order of $10^{11}$ neurons); other times it is due to the inability of our current tools to record all the relations among them (genes that express at low levels are difficult to detect); still other times it is due to other constraints (online social media companies may not release their data due to privacy concerns). We do not require a system fragment to itself operate as a system; that is, a system fragment may not necessarily have the ability to fully function or interact with its environment. Consider the complex system of cell metabolism in humans. Even with contemporary tools, we do not have access to all data pertaining to this system. In order to study it, we usually focus on a single aspect most relevant to the question at hand: for example, the set of all measurable proteins and the set of known protein complexes that they form. We refer to the combination of these two sets as the ``protein complex fragment'' of the cell metabolism system.

\item \textbf{Dependency:} a property of a system in which the existence of one relation can provide information about the existence of another relation.

\item \textbf{Formalism:} a mathematical framework which can be used to represent, model, encode, and study a complex system. In this paper, we will explicitly discuss the graph, simplicial complex, and hypergraph formalisms.

\item \textbf{Representation:} a mathematical or computational encoding of a specific complex system (or a fragment of one). A representation is the materialization of a specific formalism, e.g. it is one concrete, specific graph, as opposed to the mathematical theory, or formalism, of graphs\footnote{For readers familiar with object-oriented programming, we liken the difference between ``formalism'' and ``representation'' to that between ``class'' and ``object".}.

\item \textbf{Encode:} the process of taking a system or data collected from a system and formulating it as a representation using a specific formalism. 

\item \textbf{Attribute:} a property or bit of information attached to a node or relation. We can call the set of properties $P$ and let $p$ be the assignment map sending $V\times R \rightarrow P$. For example, a relation formed by the co-firing of neurons can be assigned a frequency, and a relation formed among individuals can have a categorical attribute such as ``teammates". In this work, we focus less on attributes and more on how we handle vertices and their relations, but we will note how node and relation attributes can extend the base formalisms. 

\end{itemize}

\section{Dependencies by the system, for the system}
\label{sec:dependencies}

When studying or modeling a complex system composed of many parts, several design decisions must be made. We begin by considering one specific and rather fundamental choice, which is sometimes only implied and other times outright neglected. This choice regards the decision of which \textit{system dependencies} one should seek to appropriately and accurately encode. Reiterating our definition above, \textit{a dependency is a property of the system in which the existence of one relation provides information about the existence of another relation.} Said another way, does the system have underlying rules or restrictions that cause interactions to occur or nodes to behave in particular ways? For example in a social system of individuals and friendships, if two individuals live physically close to one another, then their likelihood of becoming friends is larger than if they lived far apart. Furthermore, if they live near each other, then they are also more likely to meet and consequently befriend each other's neighbors. In this way, knowledge of the existence of one friendship informs us of the possible existence of other friendships, because the friendships (relations) between people (units) are affected by geographical distance (dependency). Such system-level dependencies can manifest in different ways; here we will constrain ourselves to a discussion of three of the most commonly observed dependency types. Specifically we discuss subset dependencies (does a large relation influence the existence of smaller sub-relations?), temporal dependencies (does temporal nearness of elements influence their relations?), and spatial dependencies (does the physical proximity of elements influence their relations?). We acknowledge that dependencies other than those described in this work exist within real-world systems; in many domains of inquiry, ongoing research efforts seek to define the proper avenues for illuminating dependencies and approaches for their incorporation.

\subsection{Subset dependencies}\label{sec:group_dep}

When investigating a complex system, we often record its elements and the observed relations containing two or more of those elements. For example, we might record objects and shared observable features \cite{mcrae2005semantic}, people and shared conversations \cite{zhu2017network}, or neurons and their co-firing \cite{curto2017can}. Here, we can think of the system as a set of nodes $V$ and a set of observed relations $R$ in which each relation $r \in R$ is a subset of $V$ and is meant to represent one observed interaction between $k$ elements. In this setup, some nodes may participate in many relations, while others participate in very few or none at all. It is then important to ask: if we observe the relation $r = \{v_0,\dots,v_{k-1}\} \in R$, does it imply that some subset $r'$ of $r$ is also a relation? If so, the system exhibits the type of dependency that we call a \emph{subset dependency}. For example, in the words-and-features system fragment, if three words correspond to objects that share a particular feature (so that $r = \{v_0, v_1, v_2\}$), then any two of the objects must also share that same feature (then $r' = \{v_0, v_1\}, r''=\{v_1,v_2\},$ and $r'''=\{v_0,v_2\}$ are all relations). One can make a similar argument for people conversing with one another and for neurons co-firing. In these cases, every subset of any set of related nodes is also related. However, we will see examples later when only some, or none, of the relation subsets are also relations, and we will describe this scenario as indicating the presence of a different type of dependency. Concretely, \textit{we will say that a system of nodes $V$ and relations $R$ exhibits a subset dependency if for $r \in R$ and $r'\subset r$, we must have that $r'\in R$ whenever $P(r')$ is true, where $P$ is some logical predicate.} In the case when $P$ is always true, then any subset of $r$ is always a relation, as in the examples proffered above.

\begin{figure}
    \centering
    \includegraphics{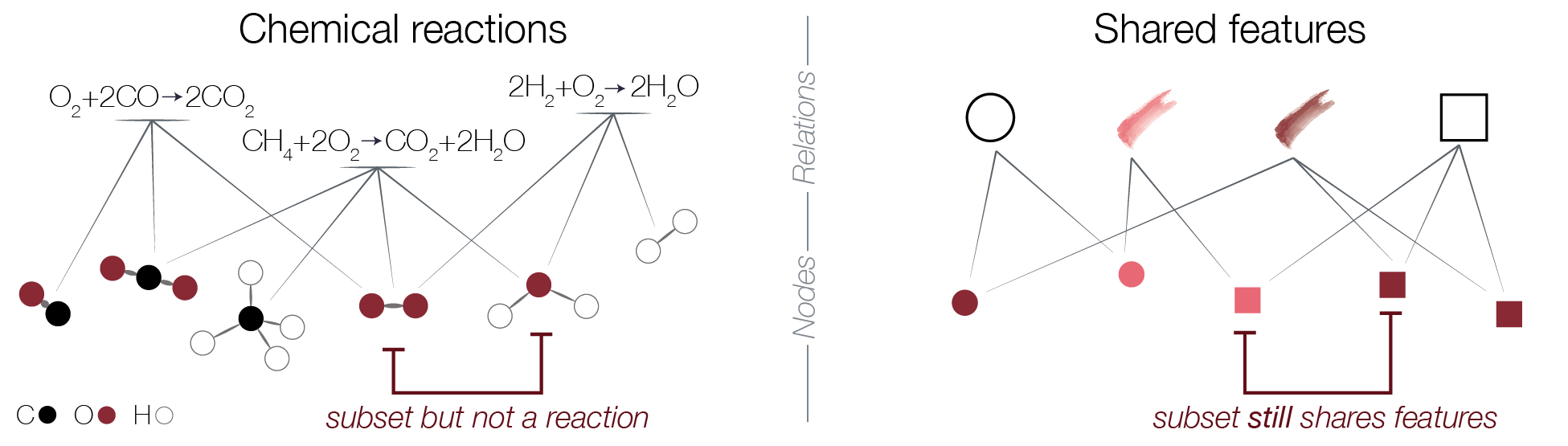}
    \caption{\textbf{Are subsets of related nodes necessarily related?} Systems may exhibit a subset dependence, which occurs when a relation between nodes implies the existence of a relation between any subset of those nodes. \emph{(Left)} An example system fragment composed of molecules and chemical reactions. Here we may have a set of molecules that all participate in the same reaction, such as $O_2$, $H_2O$, and $H_2$, but a subset of these compounds could not independently engage in a reaction, such as $O_2$ and $H_2O$. \emph{(Right)} An example system fragment composed of objects with observable features such as color and shape. All objects that are squares are related by the presence of the shared feature "square". Any subset of these square objects will also still possess the shared feature "square", and thus will also be related. Since any subset of a relation must also be a relation, we say that the system contains a \emph{subset dependency}.}
    \label{fig:rel_dep}
\end{figure}

To illustrate this specific type of dependency, in Fig.~\ref{fig:rel_dep} we show a system fragment of chemical reactions (left) and a system fragment of objects with shared physical descriptors (right). On the left side of Fig.~\ref{fig:rel_dep}, molecules or compounds correspond to nodes, and reactions define relations between nodes so that if $k$ compounds together exclusively form the reactants and products of one reaction, then those $k$ nodes are related. We see that $O_2$ and $H_2O$ participate in multiple reactions together, for example $2H_2O + O_2 \rightarrow 2H_2O$, but we do not observe a reaction that \emph{exclusively} uses $O_2$ and $H_2O$ (we would need at least one more compound using $H$ to even begin balancing that reaction). Therefore this system fragment does not display the property that subsets of relations are also relations, since we have that $\{O_2,H_2O\} \subset \{H_2,O_2,H_2O\}$ and $\{H_2,O_2,H_2O\} \in R$, but that $\{O_2,H_2O\} \not\in R$. In contrast, the right side of Fig.~\ref{fig:rel_dep} shows a collection of objects and features (shape and color), in which each object may share physical features with other objects. In this case a relation $r_{\square} = \{\mysquare{mypink}, \mysquare{myred}, \mysquare{myred}\}$ contains all objects that are square. Notice that by our definition of relation for this system fragment, we immediately get that $r' = \{\mysquare{mypink}, \mysquare{myred}\}$ is also a relation. Specifically, the pink and red squares are related because they share the feature ``square'', but also any subset of the squares will also be related because they, too, share the feature ``square''. This example of objects and shared features does display the subset dependency, since subsets of related nodes are also related.

When a system displays a subset dependency, we must ask ourselves whether we should explicitly represent that property in our model. The answer to that question will depend on, among other things, the available data, the research question, and how we define relations among nodes. Incorporating the subset dependency in a representation usually requires the data to include recorded polyadic relations, which are not always directly observable. Additionally if the research question involves paths through related nodes, it might not be necessary to incorporate polyadic relations and thus the system's subset dependencies explicitly, since often we can answer questions about paths between nodes using exclusively dyadic relations, which are simpler to compute with than polyadic relations. Most commonly, the choice of whether to include the subset dependency affects the formal representation used to encode the system, and consequently the results of downstream analyses. For example, if Marta is involved in a group of people having conversations and we define relations as shared conversations (so that a subset dependency exists), then if we count the number $p$ of people with whom Marta converses we do not know if Marta had $p$ separate conversations with each of the $p$ individuals, or if she participated in one large conversation with all $p$ people. Without a distinction, Marta's popularity with others could be vastly over- or under-estimated. This example illustrates how the occurrence of subset dependence is determined by the definition of relation. In Section~\ref{sec:representations} we explore the benefits and drawbacks of a few abstract formalisms that capture different types of dependencies. For now, we stress that the presence or absence of subset dependencies influences the computations we can perform and the formalisms we can use.

\subsection{Temporal dependencies}\label{sec:temporal_dep}

Next we consider systems in which we observe information, individuals, or goods moving along paths across time. A simple example would be a city subway system where passengers ride the train from one stop to the next until they reach their destination. In such systems we must ask the question: Does the current location of an individual affect where they might move next? \textit{We say a system exhibits a temporal dependency if the behavior of a unit at time $t$ affects the behavior of any unit at some time $t' > t$.} Said another way, paths or walks within systems that display temporal dependency are not Markovian, since the future state of a unit depends not only on its current state but also some past states of itself or other units.

\begin{figure}
    \centering
    \includegraphics{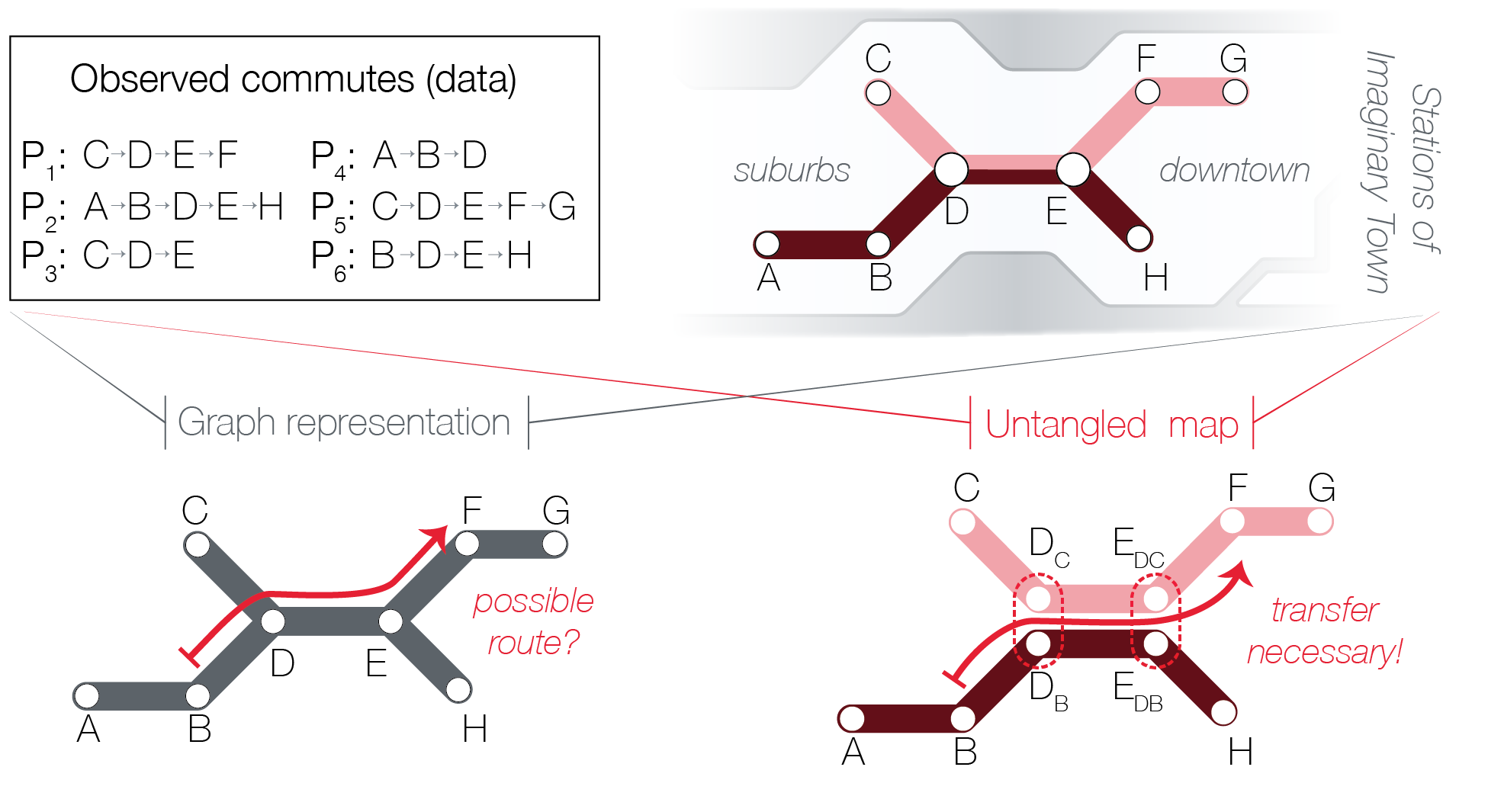}
    \caption{\textbf{By incorporating temporal dependencies into the representation, we obtain a more accurate subway map.} Given data from six commuting passengers ($P_1$, $P_2$, ..., $P_6$) who do not switch trains \emph{(top left)}, how can we obtain the underlying subway map? We could create a graph in which two stations are connected if a passenger transferred from one station to another. However, such a graph would suggest that a passenger could commute from station $B$ to $F$ without switching trains \emph{(bottom left)}, which is not possible in this system. If instead we untangle the subway lines by respecting the temporal dependency and treating trains that arrive to station $D$ from station $C$ as different from those arriving from station $B$, then we can clearly see the necessary transfer between subway lines required for the $B$ to $F$ commute \emph{(bottom right)}.}
    \label{fig:temp_dep}
\end{figure}

Consider a subway system in which trains can travel to stations $A$ through $H$ (Fig.~\ref{fig:temp_dep}). If our complex system consists of passengers commuting via the subway, then our observed data might include explicit passenger routes. For example, in Fig.~\ref{fig:temp_dep} we record the routes of six passengers, each of whom commutes from the suburbs (stations $A$, $B$, and $C$) to downtown (stations $D$, $E$, $F$, $G$, and $H$). For the purpose of the example, we assume that passengers do not transfer between distinct train lines during their commute. If we now represent our data as a set of units (stations) and we connect two units $i$ and $j$ if station $j$ immediately follows station $i$ in at least one passenger route, we obtain the subway map shown in the bottom left of Fig.~\ref{fig:temp_dep}. Because this diagram records all known movements of passengers between pairs of stations, we might confidently proceed to the next analysis step. However, it is worth noting that this particular representation suggests that the green path from station $B$ to station $F$ is a possible commute for a passenger. Yet when we look back at the data itself, such a commute seems extremely unlikely since the sequence $B-D-E-F$ never occurs. The fact that this route appeared natural from the representation, but not from the data, points to the fact that our system contains a temporal dependency and, importantly, that this dependency is not well reflected in the particular representation we chose.

As discussed in great detail in \cite{benson2016higher,rosvall2014memory,edler2017mapping,de2015identifying,lambiotte2014effect,perri2019higher}, the fundamental limitation of keeping only pairwise sequential relations, as done in the bottom left of Figure~\ref{fig:temp_dep}, is that in the representation we assume that traversal across each link is Markovian and therefore its probability is independent of the probability of traversing any other link in the system. More explicitly, paraphrased from \cite{lambiotte2019networks}, by representing the system as a graph (see Section \ref{sec:representations} for a definition) we assume that the edges $(i,j)$ and $(j,k)$ are independent and that the two-step transition from $i$ to $k$ proceeds in two independent steps. This assumption can easily be violated by a real system, as seen in our toy example, since sometimes one step in this traversal is dependent on which steps came before (i.e. transitions are not Markovian). Mismanaging temporal dependencies in systems can lead to biased results that can, for example, over-represent the importance of edges rarely used or create non-existent connections (as in the case of our subway example). We will discuss a formalism particularly appropriate for representing temporal dependencies in Section~\ref{sec:representations}.


\subsection{Spatial dependencies}\label{sec:spatial_dep}

The third and final type of dependency that we discuss here arises from the physical nearness of units within a system. For example, in the human connectome a brain region is likely to extend white matter tracts to neighboring regions, providing physical conduits for electrical activity \cite{stiso2019white}. In granular materials, resistance to external forces relies on interactions between only particles that physically touch \cite{papadopoulos2018particles}. More generally, many spatial systems are so named because the spatial location of nodes affects their likelihood of interacting with one another \cite{barthelemy2011spatial,barthelemy2018transitions}. \textit{Here we say that a system exhibits a spatial dependency if the distance between two or more nodes influences the existence of a relation that contains them.} Typically, this dependency is encoded by enforcing that the probability of the occurrence of a relation in a concrete representation is a function of the distance between the nodes involved.

\begin{figure}
    \centering
    \includegraphics{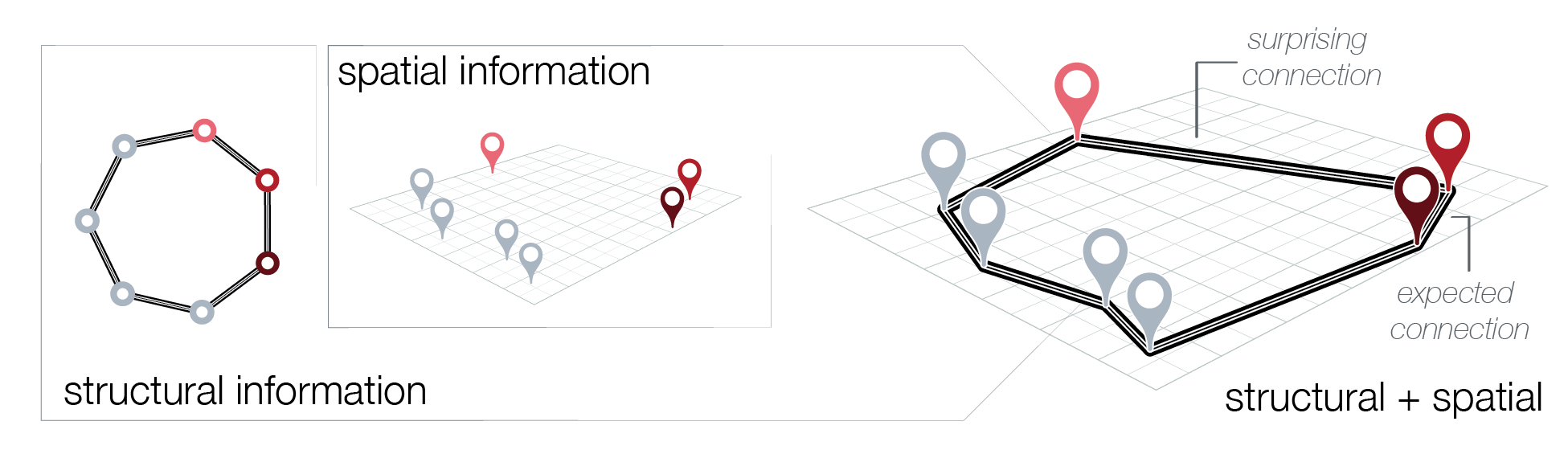}
    \caption{\textbf{Spatial dependencies within a system can complicate our representations of the data.} In our example system, we have \emph{(Left)} connection information that is independent of any system embedding, and \emph{(Middle)} spatial information indicating where the nodes physically reside. A possible combination of the two information types \emph{(Right)} can be used to better understand the physical constraints on the topology. If long distance connections are costly for that system, the combined representation allows the investigator to assess the prevalence and location of those costly (and thus potentially surprising) connections.}
    \label{fig:spat_dep}
\end{figure}

Many such systems exist in the natural and manufactured world. Indeed, spatial restrictions influence communication in cell populations \cite{lai2004notch,ramel2013rab11}, trade in economic networks \cite{hu2013research}, and passengers in transportation networks \cite{weber2002bringing,lima2016understanding}. As an example of spatial dependency within an abstract system, we might begin with only knowledge of the pattern of related nodes. We display this structural information in the left panel of Fig.~\ref{fig:spat_dep} with circles corresponding to nodes and lines joining circle pairs whose corresponding nodes are related. From the structural information alone we might expect that relating the pink and red nodes is just as difficult or costly as relating the red and dark red nodes; we might therefore infer that the two relations are equally crucial to the system's function. However, if the system exists within an environment containing coordinates and a distance function, with each node having spatial coordinates and a measure of distance between each pair, then this spatial information could offer a different perspective on the system. In the middle panel of Fig.~\ref{fig:spat_dep}, we see that the nodes, now depicted with colored pins, are spread out so that some are more spatially clustered whereas others are less so. Considered alone, the spatial information gives us no insight into the actual relations present in the system, but does provide information with which we might predict the likelihood that nodes are related. 

In many spatial systems such as the brain or city transportation, relations between distant nodes are unfavorable due to a higher cost of creation and maintenance, while short-range relations are far easier to construct. In the face of this association between the physical distance across a relation and its cost, we might consider the distances between nodes and infer that the red and dark red nodes are likely to be related, while the pink and red nodes are not. When we finally combine the topological and spatial information (Fig.~\ref{fig:spat_dep}, right), we then can leverage the two information types to understand which relations are most surprising or make hypotheses about which relations are most important to the system. For example, the dyadic relation between the pink and red nodes might be very costly given the long distance, so we might infer that the pink to red relation is more essential to the system than the red to dark red relation since the system would only spend valuable resources to maintain such a relation if it was integral to system function. Without the spatial information, we may have incorrectly placed the same importance on the pink-to-red and the red-to-dark red relations. This example highlights one of many ways in which we could integrate spatial and topological information.

As with the previous dependency types, failure to account for a spatial dependency can greatly bias our models and results. Consider an outbreak of a contagious disease. If we recorded the habits of infected individuals such as their diet, but fail to record their locations and physical mobility through space \cite{tizzoni2014on,apolloni2013age}, then we might -- for example -- wrongly attribute disease spread to the broad consumption of a particular food that is prevalent in the infected region instead of through person-to-person contact. As another example, social contacts are also influenced by proximity. If we return to evaluating Marta's popularity, the observation that she has many friends may come from the fact that she lives in a densely populated area, rather than from her charisma or personality. In these examples, failing to account for spatial dependencies may result in attributing certain structural properties of the system to the wrong cause.

\subsection{External sources of dependencies}\label{sec:data}

Before we shift our focus to concrete ways of encoding system dependencies using mathematical formalisms (Section~\ref{sec:representations}), it is useful and interesting to consider how external forces can influence the observed system dependencies. Ideally, we as investigators would have the ability to measure all dependencies within the system under study, and then use this knowledge to make an informed decision as to the appropriate formalism with which to model our system. However, often the processes of scientific inquiry do not proceed so effortlessly: no analysis is ever devoid of the influence of external factors, or biases. Our goal in this section is to highlight possible sources of such bias. Although we have already discussed biases arising from dependencies native to the system under study, here we emphasize that acknowledging and understanding dependencies imposed by outside sources should also play a crucial role in determining an appropriate representation and subsequent analyses.

\begin{itemize}
\item \textbf{Data availability.} One notable and common constraint in science is the limited data that can be empirically acquired from a given system, i.e. researchers usually have access only to a fragment of the system. As a consequence, any subset dependency that is observed and ultimately encoded may be determined more by the sparsity of available data than by the system's true structure and function. For example, one may have access to only sparse snapshots of or short sequences from an evolving system \cite{sinatra2010networks}, making the subset dependency difficult to identify and effectively encode. Particularly, there may not be enough data available to correctly deduce the predicates $P$ that a subset must satisfy in order to also form a relation (see the definition of subset dependency in Section \ref{sec:group_dep}). 

\item \textbf{Data acquisition or processing.} Certain experimental techniques or computational procedures may produce spurious dependencies. A common example involves correlation matrices. By computing the correlations of node activity one induces a transitivity dependency, which is a type of subset dependency. Concretely, if $A,B,C$ are nodes in a system where two nodes are related if the time series of their activities are highly correlated to each other, as determined by some data acquisition method, then whenever $A$ and $B$ are related, and $B$ and $C$ are related, it is highly likely that $A$ and $C$ are also related. In this case, it is possible that relations between nodes implied by the calculated correlations are found in the processed data but not in the system itself. For example, one might find that changing the type of correlation (or other similarity measure) results in a change in the inferred node relations.

\item \textbf{Research question.} The research question at hand will influence which relations within a system are particularly interesting. Moreover, it may also influence the very definition of a relation. For example, consider a system of proteins that interact to form protein complexes. If we wish to study which proteins appear together in many complexes, then we may define a relation as two proteins that participate in the same complex. If instead we wish to study protein complexes themselves, we could define a relation as a set of proteins that all together form a single complex. In the first case, the relations exhibit a subset dependency (if three proteins appear together in a complex, then so do any two of them), but the second does not. On the flip side, a given research question may neglect a relevant dependency in the system. For example, we could ask if a common food could have caused a disease outbreak. Answering that explicit question neglects the fact that individuals near each other will likely eat similar foods. The research question is not broad enough to incorporate the spatial information as part of the answer, and therefore spatial dependencies may seem irrelevant at first sight, when they may be in fact essential to finding the real answer. We expand upon this topic in Section~\ref{sec:multiple_dep}.

\end{itemize}

\begin{figure}
    \centering
    \includegraphics[width = \textwidth]{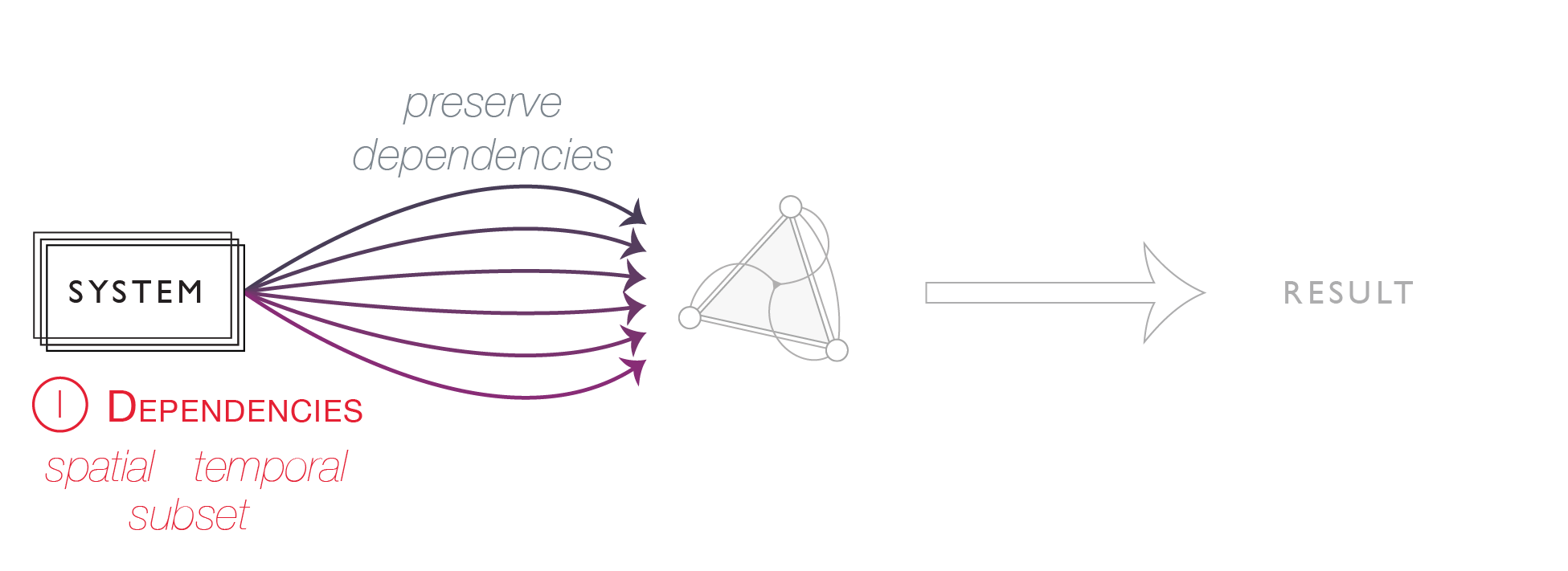}
\caption{\textbf{Understanding dependencies present in the system is a first step in the complex system analysis pipeline.} Types of dependencies include spatial, temporal, and subset dependencies. Acknowledging dependencies at this step allows for proper preservation of dependencies throughout the rest of the analysis pipeline.}
    \label{fig:flow_deps}
\end{figure}

To summarize, we have defined and discussed three types of dependencies that could exist in a complex system: subset, temporal, and spatial. We emphasize that dependencies can arise from within the system itself or from external factors, but regardless of their origin, we as researchers must be aware of their existence and how they influence our models and results, especially given their early position in our analysis pipeline (Fig.~\ref{fig:flow_deps}). As we will continue to see in the sections that follow, the recognition and encoding of dependencies can greatly affect the results of our analyses and the conclusions that can be drawn.

\section{Formal representations of complex systems}\label{sec:representations}
Over the years many representations of complex systems coming from different mathematical and computational formalisms have taken hold across scientific disciplines. Different formalisms allow for the modeling of unique aspects and dependencies of each system, but the multiplicity of available formalisms presents challenges for the communication, collaboration, and ultimately the progress of complexity science. Furthermore, the choice of formalism also complicates the analysis pipeline that researchers must decide upon when studying a particular system.

Here we discuss three of the many possible mathematical formalisms that researchers commonly use to represent their system: graphs, simplicial complexes, and hypergraphs, chosen for their prevalence in the complex systems literature. A complex system is, at its core, a collection of units and their relations, so therefore we require our representations to mirror this composition of units and relations. The units of all three formalisms discussed here are called \emph{nodes}. \emph{Graphs} represent pairwise relations among nodes as \emph{edges}. Despite their simplicity (or perhaps because of it), graph representations have supported several important discoveries such as the prevalence of small-worldness in real-world networks \cite{watts1998collective,amaral2000classes}. Still, graphs can only, by nature, represent dyadic relations between nodes. If instead relations within the system exist between more than two nodes, one might turn to either a \emph{simplicial complex} or a \emph{hypergraph}. Both of these formalisms naturally allow us to encode such polyadic relations \cite{battiston2020beyond}. The relations represented by a simplicial complex are called \emph{simplices} and those represented by a hypergraph are called \emph{hyperedges}. We will first define each formalism, so that later in this exposition we can explicitly discuss their respective advantages and assumptions.

\subsection{Graphs}

\begin{figure}
    \centering
    \includegraphics[width = \textwidth]{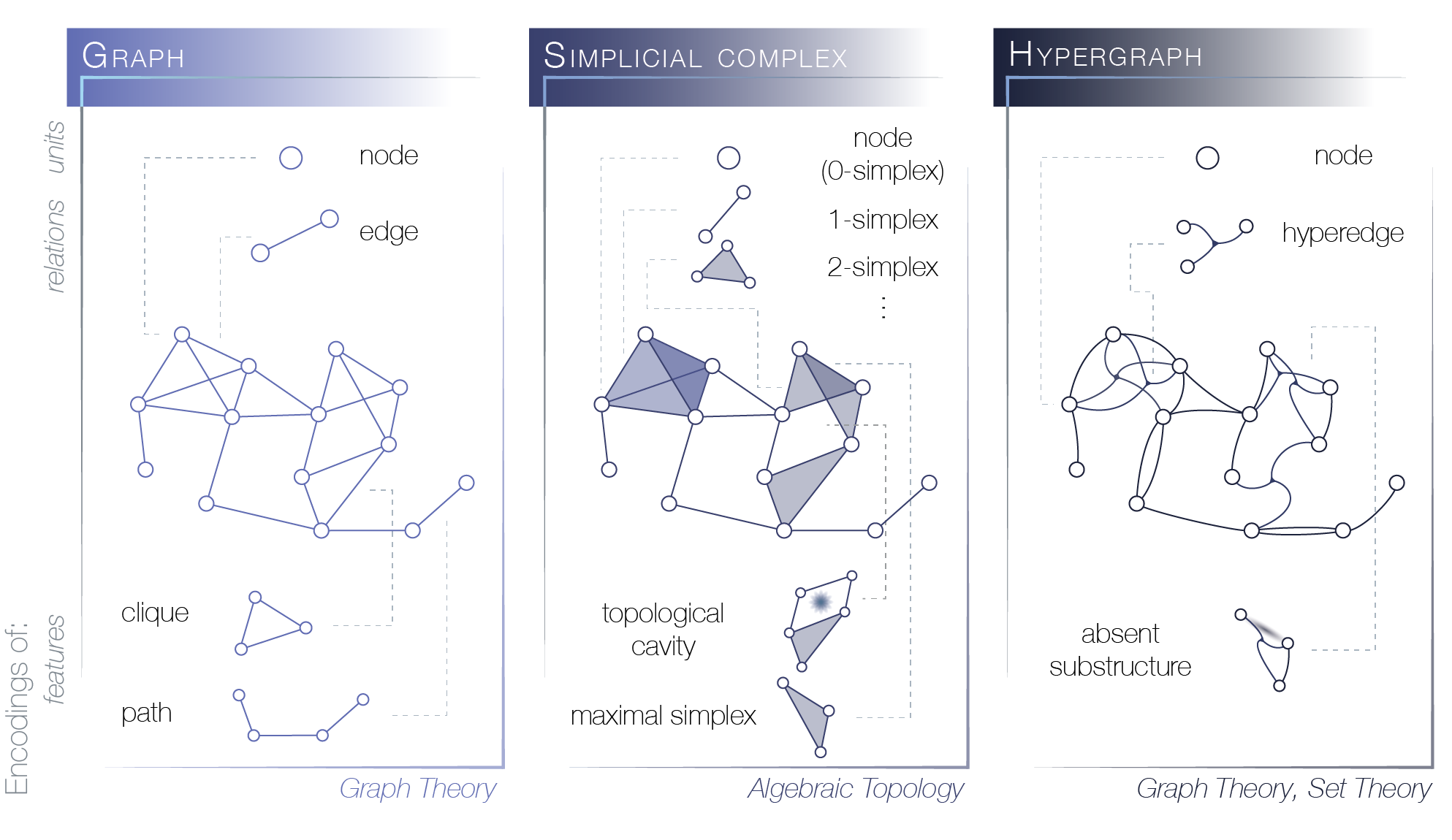}
    \caption{\textbf{Three types of formalisms composed from nodes and relations}. \emph{(Left)} Graphs involve units called nodes and relations between two nodes called edges. Possible features of interest for graphs include all-to-all connected sets of nodes called cliques, as well as routes between nodes called paths. \emph{(Middle)} Simplicial complexes can be used to represent systems with polyadic relations among units. Sets of related nodes are connected by simplices. A $k$-simplex describes $k+1$ nodes that collectively interact, such that any subset of nodes forming a simplex must also form a simplex; this is called ``downward inclusion''. Motifs of interest include topological cavities and maximal simplices. \emph{(Right)} Hypergraphs can also be used to represent systems with polyadic relations among units. Sets of related nodes are connected by hyperedges. Hypergraphs are not restricted by downward inclusion.
    Of particular interest within a hypergraph is the absence of a substructure (or smaller relation), for example in which two nodes do not connect dyadically but participate together in a hyperedge that connects a superset of the node pair.}
    \label{fig:graphs}
\end{figure}

The first and perhaps most common formalism used to model complex systems stems from graph theory. A \emph{graph} $G$ is a collection of vertices and edges between vertices such that an edge connects exactly two vertices (Fig.~\ref{fig:graphs}, left). We denote the set of vertices as $V$ and the set of edges $E\subseteq V \times V$, so that a graph is defined uniquely by $G=(V,E)$; note that each edge is an unordered set of two nodes. The vertices of a graph are the main units, and edges describe how these units fit together. If $v_A$ and $v_B$ are nodes of the graph, then we write $(v_A, v_B)$, or $v_A - v_B$ to represent the fact that the two nodes are connected by an edge. Studies that form a graph representation from the underlying data frequently involve finding densely connected sets of nodes or determining how an object might traverse the structure. In using the graph representation, such questions could lead to detecting cliques or communities in the graph, or identifying chains of connected nodes called paths in the graph (see Section~\ref{sec:methods-for-graphs} for more examples).

Many attribute the origin of graph theory to Leonhard Euler in the 18th century \cite{euler1741solutio}. One can also trace its presence outside of mathematics back to the use of sociograms and social network analysis in the 1930s \cite{freeman2004development}, and to graph-like data structures in computer science in the 1950s \cite{wilson2013history}. Notably, the use of graphs to model more general complex systems has rapidly increased over the past few decades, driven largely by the discovery of the small-world effect \cite{watts1998collective} and heavy-tail degree distributions \cite{barabasi1999emergence} in real-world datasets. Encoding a system as a graph has the great advantage of hundreds of years of mathematical theory behind concepts, generally simple computations, and insightful visualization. However, the graph by definition assumes that relations between nodes occur exclusively at the pairwise level. Systems such as transportation networks might solely contain pairwise relations, but many others, especially from biology, often have polyadic relations. Still, the graph's ability to model systems has proven quite useful in distinct fields such as neuroscience \cite{bassett2017network,bullmore2009complex}, computer science \cite{even2011graph,mitchell2006complex}, and ecology \cite{proulx2005network,montoya2006ecological}.

\subsection{Simplicial Complexes}

The next formalism that we consider addresses the need to acknowledge polyadic relations in the system. A \textit{simplicial complex} is a set of vertices $V$ along with a collection of subsets of vertices $R$ (our set of relations, often denoted by $K$ in the field) such that for any $r \in R$ and $r' \subset r$, we have $r' \in R$ (see Fig.~\ref{fig:graphs}, middle); we will refer to this condition as ``downward closure''. A set of $k+1$ vertices $r \in R$ is here called a $k$-simplex, and downward closure requires that any subset of vertices within a simplex also forms a simplex. In practice we often imagine a $k$-simplex to indicate an application-relevant interaction between the $k+1$ nodes, such that these nodes may function in unison. The simplicial complex (precisely, the \textit{abstract simplicial complex}) would then record the individual units (nodes), the functional building blocks (simplices), and how all these building blocks are assembled into one system (the simplicial complex). Since subsets of simplices are simplices by definition, the natural intuitions of node relations apply readily; that is, if $k$ nodes are related, then any subset of those $k$ nodes are also related. The simplicial complex can be easily written as an $\#maximal$ $simplices \times \#vertices$ binary matrix where an element containing a 1 indicates vertex participation in the given maximal simplex, where a maximal simplex is a simplex that is not contained in any larger simplex. 

Although algebraic topology has been studied for well over a century, it was not until the early 2000's that applied algebraic topology as a discipline began to emerge \cite{zomorodian2005computing,edelsbrunner2000topological} (though we note a few earlier uses \cite{atkin1972cohomology}). Many of the earliest studies used applied topology and simplicial complexes to study data in the form of point clouds \cite{carlsson2008local,singh2008topological}. Later, it became clear that the simplicial complex language was a natural formalism for explicitly representing biological and physical systems. For example, simplicial complexes have been used to represent neural recordings \cite{giusti2015clique,curto2017can}, classify images \cite{tauzin2020giotto,damiano2018topological,dunaeva2016classification}, and describe the mesoscale architecture of brain networks \cite{stolz2014computational,stolz2018topological,reimann2017cliques,sizemore2018cliques, petri2014homological}. Even more recent work has focused on defining generative models to construct simplicial complexes with given topological features \cite{courtney2018dense}.

\subsection{Hypergraphs} 

The final formalism that we consider draws again from sets of nodes and their relations, yet is even more general than the simplicial complex discussed above. The \emph{hypergraph} is an extension of the mathematical definition of a graph, in which we have a vertex set $V$ and a hyperedge set $\mathscr{E}$. A hyperedge $e\in \mathscr{E}$ (using our notation this phrase reads a set $r \in R$) can connect an arbitrary number of vertices. That is, while an edge can only connect two vertices, a hyperedge can bridge three, four, five, or more nodes (Fig.~\ref{fig:graphs}, right). More rigorously, a hypergraph is a pair $(V,\mathscr{E})$ with $V$ a finite vertex set and $\mathscr{E}$ a set of subsets of $V$ \cite{voloshin2009introduction,berge1984hypergraphs}. In contrast to the simplicial complex, we can use the hypergraph to encode polyadic relations without the restriction of downward inclusion. Formally, a subset $e'$ of a hyperedge $e$, $e' \subset e\in \mathscr{E}$, does not necessarily exist as a hyperedge. Additionally, we can rewrite a hypergraph as a $\times \#hyperedges \times \#vertices$ binary matrix, in which an entry of 1 indicates the vertex participation in the hyperedge. 

As noted above, the crucial restriction that is relaxed when moving from describing a simplicial complex to defining a hypergraph is that of downward closure. Recall that in a simplicial complex, if $r$ is in the simplicial complex, any subset $r' \subseteq r$ must also be in the simplicial complex. Hypergraphs do not obey this rule. For example we may see a hyperedge connecting vertices $v_1,v_2$, and $v_3$ but no hyperedge that connects $v_1$ to $v_2$ exclusively. Or, given two hyperedges connecting nodes $v_1, v_2, v_3$, and $v_2, v_3, v_4$, if a hyperedge connecting $v_2,v_3$ also existed, does this smaller hyperedge indicate a sub-relation for the hyperedge $v_1, v_2, v_3$, the hyperedge between $v_2,v_3, v_4$, neither, or both? With a hypergraph, we cannot determine how a sub-relation fits (or does not fit) into superset relations (see \cite{spivak2009higher} for a deeper discussion). This subtle difference allows hypergraphs to represent a wide diversity of systems, including many that the simplicial complex formalism would not appropriately represent. The hypergraph's increase in modeling flexibility is counterbalanced by a decrease in formal analysis methods, which we will discuss more in Section \ref{sec:methods}.

The flexibility and ability to model polyadic relations made hypergraphs an appealing formalism in many systems that were originally studied with graph theory. Indeed one of the earliest practical uses of hypergraphs was to understand social networks \cite{seidman1981structures}. Since then, researchers have successfully employed hypergraphs to study polyadic relations in the Enron email dataset \cite{purvine2018topological}, find the core of yeast protein-protein interactions \cite{ramadan2004hypergraph}, uncover motifs in neurodevelopment \cite{gu2017functional}, track changes in evolving systems \cite{bassett2014cross,davison2015brain,davison2016individual}, and detect failure in biochemical networks \cite{klamt2004minimal}. As many uses of hypergraphs arose out of systems first modeled with graphs, many analysis methods for hypergraphs mimic those originally used for graphs (we discuss this point further in Section \ref{sec:methods-for-hypergraphs}).

\subsection{Variations}\label{sec:variations}

We note that the above descriptions only scratch the surface of complex system encoding possibilities. An ever broadening set of scientific questions drives the need for novel variations of each formalism, resulting in a myriad of definitions and manipulable parameters. One could extend our mathematical definition of complex systems to include the following properties, perhaps as a map $p:V \times R \rightarrow P$ where $P$ is a set of attributes we care about, as mentioned in Section \ref{sec:definitions}. Here we note a few of the most common modifications to each of the above formalisms, driven by the need to incorporate more information about the system at hand.

\subsubsection*{Directed}

Many complex systems including the brain, transportation networks, and metabolic pathways exhibit directionality in their relations. That is, in these systems, if $v_A$ and $v_B$ are units that share a dyadic relation, there is a meaningful distinction between a relation where $v_A$ comes first, one where $v_B$ comes first, and one where either $v_A$ or $v_B$ comes first (but there must always be an order in how they are related). To distinguish these cases we write $v_A\rightarrow v_B$, $v_B\rightarrow v_A$, or $v_A\leftrightarrow v_B$, respectively. If we apply this idea to the graph formalism, a \emph{directed graph} is one where each edge is now an ordered set of two nodes. Directed graphs have proven extremely useful in many contexts from scheduling and monitoring workflows \cite{kotliar2019cwl,apache_airflow} to cardiac excitation modeling \cite{vandersickel2019directed} to understanding percolation processes relevant to wild fires and other explosive phenomena \cite{squires2013weakly,raissa2019explosive}. Moving to simplicial complexes, directionality is still quite natural. Indeed simplices themselves inherit a directionality, formally known as an \emph{orientation}, encoded by the natural numbering of the participating vertices. In practice, an oriented $k$-simplex implies that the $k+1$ vertices, and any subset of these vertices, all relate to one another such that we could number the vertices so as to ensure that vertices only point to vertices with a higher assigned number. Oriented simplicial complexes arise in practice from directed synapses between neurons \cite{reimann2017cliques} as well as directed migration flow \cite{ignacio2019tracing}. Finally, in hypergraphs, one may represent directionality with \emph{hyperarcs}, the term for a directed hyperedge. More formally, a hyperarc is a pair of disjoint subsets of vertices with one subset comprising the sources and the other subset comprising the sinks \cite{gallo1993directed}. Directed hypergraphs have proven useful in constructing a biological pathway database \cite{krishnamurthy2003pathways}, tackling problems in computer science such as propositional logic \cite{gallo1993directed} and combinatorial optimization \cite{levi1976generalized,gnesi1981dynamic}, and finding specific patterns of connectivity in chemical reaction systems \cite{ozturan2008finding}, among others.

\subsubsection*{Weighted}

Not all relations are created equal; even within the same system, relations between individual actors are rarely uniform in real-world systems. To represent these differences, the strength or magnitude of interactions between units can be encoded using the \emph{weighted} versions of the above formalisms. To weight any of the above encodings, we can define a general weight function $W:R \rightarrow \mathbb{R}$ from the set of encoded relations $R$ (edges, simplices, or hyperedges) to the real numbers $\mathbb{R}$. For a graph, this function would assign a value to each edge, which we generally interpret as the strength or frequency of the pairwise interactions between the corresponding nodes. In the context of weighted representations, the original versions containing no weights are called \emph{binary} or \emph{unweighted}, as they can be cast as weighted objects where the weights of all relations are either one, if they exist, or zero if they do not exist. The brain connectome, traffic between municipalities \cite{de2007structure}, and functional similarity of genes \cite{pan2018interrogation} have all been modeled as weighted graphs. Additionally, many common graph metrics such as the clustering coefficient and path length (covered in more detail in the next section), extend easily to the case of weighted graphs \cite{rubinov2010complex}, making this variant of representation particularly pervasive. Similarly we can construct a weighted simplicial complex by assigning a weight to each simplex. However, recall that in a simplicial complex any face of a simplex must also be a simplex, and thus if we have a relation between $k$ nodes then any subset of these nodes must be related to at least the same extent as the superset. Said another way, we require that the weighting function $W$ on simplices adheres to the rule that for any simplex $r$, if $r' \subseteq r$ then $W(r) \leq W(r')$. Weighted simplicial complexes naturally arise from point clouds with inverse distances between points as weights or growing processes. Perhaps most often, we study weighted simplicial complexes through the lens of persistent homology, which returns the organization of topological cavities housed within the weighted simplicial complex \cite{zomorodian2005computing,carlsson2009topology,ghrist2008barcodes,otter2017roadmap} (see a few recent uses in \cite{sizemore2018cliques,giusti2015clique,petri2014homological,stolz2014computational}). Lastly, in hypergraphs we can naturally weight hyperedges with distinct values \cite{gallo1993directed}. Importantly, weighting hyperedges allows more flexibility in choosing weights, as weighted hypergraphs do not enforce rules restricting weights on subedges in contrast to weighted simplicial complexes. Weighted hypergraphs have proven useful in image segmentation \cite{rital2005weighted} and in the process of incorporating prior knowledge into learning algorithms \cite{tian2009hypergraph}.

\subsubsection*{Dynamic}

Complex systems such as cell signaling, traffic patterns, and transactional relations also grow, separate, or fluctuate in time \cite{maheshwari2019model,saadatpour2012discrete,chodrow2016demand,liang2018evolutionary}. Consequently, formalisms have been adapted to represent such an evolving architecture. A \emph{dynamic graph} or a \emph{temporal graph} is a sequence of graphs $G_1,\dots, G_T$ in which each $G_i$ is a graph on the same set of nodes, and each node is mapped to its identity when moving from $G_i$ to $G_{i+1}$ \cite{holme2012temporal}. As with other variations on graphs, multiple computational tools such as community detection have been extended to include dynamics \cite{nicosia2013graph,sizemore2018dynamic,mucha2010community}. Moving to simplicial complexes, a dynamic simplicial complex is similarly a sequence of simplicial complexes on the same vertex set. Questions about the topological cavities of simplicial complexes can still be asked by using vineyards \cite{yoo2016topological} and zig-zag persistent homology \cite{milosavljevic2011zigzag} to expose the evolving topology of special types of evolving simplicial complexes. Finally, a dynamic hypergraph is a sequence of hypergraphs $H_1, \dots, H_T$ on the same vertex set where hyperedges may change from $H_i$ to $H_{i+1}$. At the time of writing, we found few examples of applied dynamic hypergraphs, although we note that their visualizations have been studied \cite{valdivia2017hypenet}. Nevertheless, we suggest that this particular variation of hypergraphs could be useful for example in modeling evolving gene interactions, functional relations between brain regions, and the time-varying structure of social groups.

\subsubsection*{Multilayer}

Often the pieces or relations between pieces of a system have types, categories, or classifications that distinguish them. It is sometimes useful to distinguish between these types of relations in our representations, and one way to do so is to use the so-called \emph{multilayer} variations. Generally, multilayer graphs consist of a set of graphs that may (or may not) involve the same nodes; each graph in the set comprises a \emph{layer}. The graph in a given layer contains relations of exactly one type. Consider a human brain in which two regions might show an increase in blood flow due to coupled neuronal activity and due to interactions involving nearby blood vessels themselves. To encode these two types of relations in a single representation, we could use a multilayer graph with two layers: one encoding the relations between neurons and another encoding relations between blood vessels. We note that when all layers contain the same set of nodes, the representation is called a \emph{multiplex} graph. We invite the interested reader to reference \cite{kivela2014multilayer,bianconi2018multilayer} for more rigorous definitions, and \cite{menichetti2016control,brummitt2012suppressing,zhong2017linear} for implications for diffusion and control. Dynamic systems can be seen as a subtype of multilayer systems, in which the layers are a set of graphs ordered in time. Previous studies have used multilayer networks to model complex spreading processes \cite{de2016physics,sahneh2014competitive,salehi2015spreading}, understand explosive word learning \cite{stella2018multiplex}, and uncover the community structure of trade relations \cite{barigozzi2011identifying}. Multilayer simplicial complexes or hypergraphs would similarly include a set of simplicial complexes (respectively, hypergraphs) not necessarily defined on the same nodes in each layer. As of the time of this writing, we did not find applications yet of this extension. We suggest that these variations could prove useful for understanding multiple types of biological data collected on a set of nodes. As an example, one could encode common properties (mutation status, chromatin rearrangements, etc.) as layers in a multiplex network of cancer cell lines in order to better understand drug response \cite{rees2019computational}). The multilayer variation is readily applicable whenever researchers have access to and want to model two different fragments of the same system.

\subsubsection*{Higher Order Networks}\label{sec:hons}
Higher Order Networks (HONs) are a variation of the graph formalism that aims to represent a certain kind of temporal polyadic relations. Instead of encoding system units as nodes, the HON encodes frequent paths or transitions in the data as nodes, which then allows us to interpret the final representation with the standard Markovian assumptions on edge sequences. Recall our example of commuting passengers in Figure \ref{fig:temp_dep}. We can build a HON from the observed path data to encode the observed dynamics and temporal dependencies of this system in a particular kind of graph. In Figure \ref{fig:temp_dep}, the more accurate subway map on the bottom right, reconstructed from the observed data, contains two nodes that correspond to the physical station $D$. The one labeled $D_B$ represents the passengers that arrive to $D$ from station $B$, while $D_C$ corresponds to those that arrive from station $C$. Similarly, the physical station $E$ splits into two nodes: $E_{DC}$ and $E_{DB}$. The nodes on this map do not correspond to the stations observed in the town's transportation system, but to the possible passenger pathways through them. Indeed, as observed before, we never observe a passenger commute that traces the path $C-D-E-H$: all passengers that pass through stations $C-D-E$, in that order, then go on to station $F$, while all passengers that pass through stations $B-D-E$, in that order, go on to station $H$. Therefore, the representation on the bottom right, an example of a higher-order network or HON, is a more faithful representation of the observed data and its temporal dependency. Note that if the observed passenger data changed to include a route visiting stations $C-D-E-H$, the structure of the HON would change, even if the physical brick-and-mortar subway system, and its graph representation, would not. We discuss HONs in the next subsection and refer the interested reader to \cite{benson2016higher,rosvall2014memory,edler2017mapping,de2015identifying,lambiotte2014effect,perri2019higher} for further details.

\subsubsection*{Further variations}

We note the above variations on the three main formalisms discussed are only the beginnings of possible ways to extend these representations. Depending on the complex system and questions at hand, certainly one may combine the variations described above to make, for example, an edge-weighted dynamic network \cite{khambhati2018modeling}, a directed multilayer network, or another combination that provides an effective representation. One may also study systems of weighted nodes instead of weighted edges \cite{sizemore2018knowledge,murphy2016explicitly}, or representations where each node has some kind of internal structure \cite{colizza2008epidemic,estrada2018metaplex}. Any of the formalisms above could also lend itself to studying the intricacies of coupled dynamical systems such as coupled oscillators \cite{papadopoulos2017development,ott2008echo} or interacting threshold-linear models \cite{morrison2019predicting}. Indeed when including variations on the three formalisms covered in this review, we find we can encode an impressive range of complex system types and properties.

\subsubsection*{Other Formalisms}

We recognize that many other formalisms intended for complex systems exist and that those we specifically mention in this review constitute only a small subset of the possibilities. Other possible formalisms include \emph{graphons}, which describe limits of sequences of graphs and can be used to estimate large, noisy systems \cite{borgs2017graphons}, \emph{metapopulation models} which classically describe global behavior of many local species populations \cite{levins1969some,taylor2011metapopulation,hanski1999metapopulation} and can be adapted to networks \cite{colizza2008epidemic}, random sequences of sets \cite{benson2018sequences}, and \textit{sheaves} which can handle added information on each node in a network and have previously been used to frame the network coding problem \cite{ghrist2011applications} and find consensus in sensor networks \cite{curry_2014}.


\subsection{Encoding system dependencies}\label{sec:multiple_dep}

As we discuss above, the formalism used to encode our data should be carefully chosen to respect any prominent properties of the system, and specifically the dependencies found therein. In this subsection we discuss the subtleties of choosing an appropriate formalism, and then review the common practices that researchers use to encode subset, spatial, and temporal dependencies using the formalisms we have introduced.

\begin{figure}
    \centering
    \includegraphics{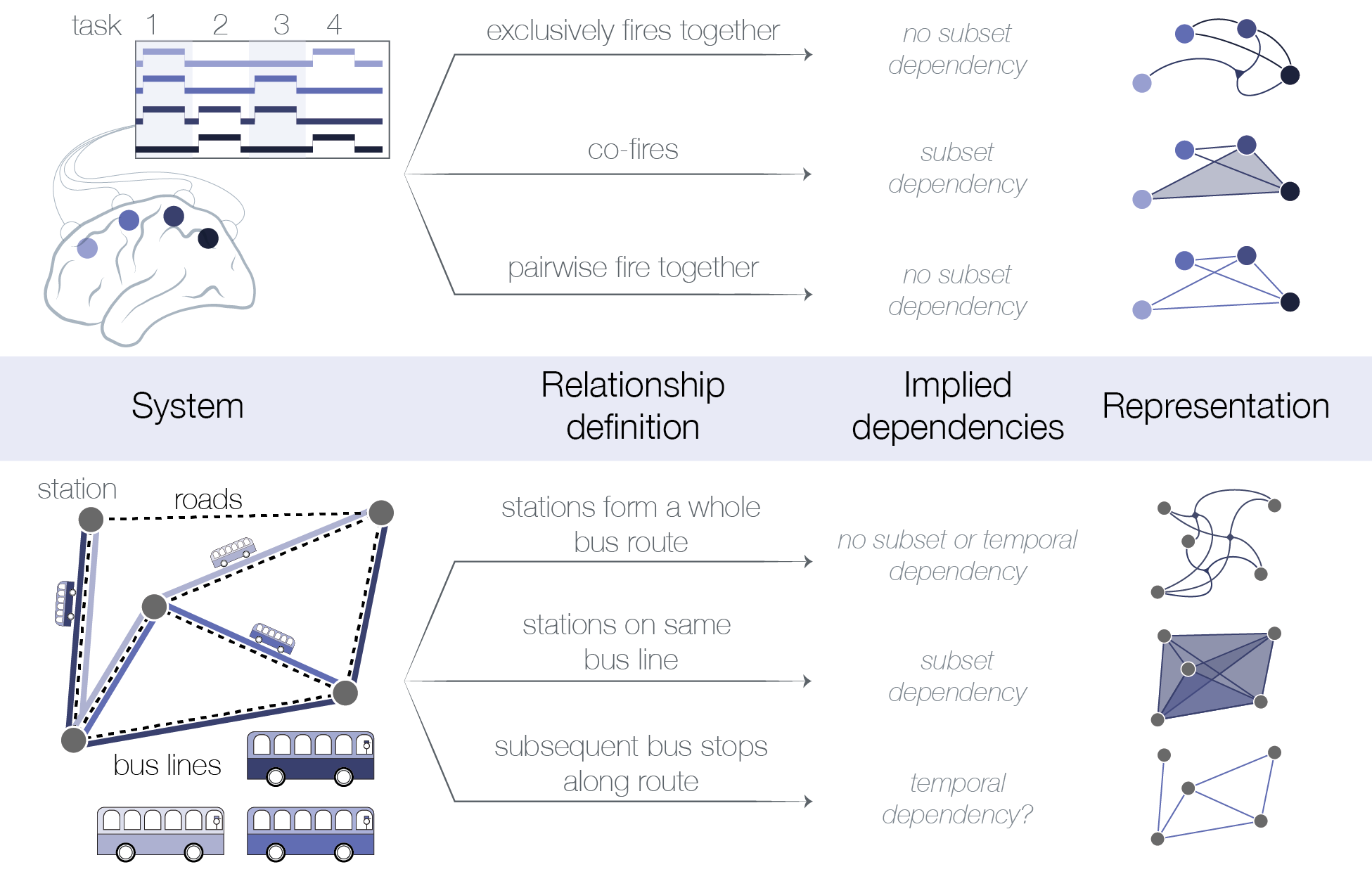}
    \caption{\textbf{Native dependencies captured by the definition of relation.} \emph{(Top)} We might record the on/off activity of four brain regions in each of four tasks (left). Depending on the definition of relation chosen (middle), we may or may not record a dependency in a representation (right). \emph{(Bottom)} Given five bus stations placed along a set of roads (dashed lines), we observe three bus lines that connect the stations (left). Depending on the definition of relation chosen (middle) we might include a subset or temporal dependency, which we would want to capture in our representation of the system (right).}
    \label{fig:relationships_deps}
\end{figure}

Once we have chosen which dependencies to model, it is important to carefully determine when two or more units in our system are related to each another -- i.e. to define the relations in our model (Fig. \ref{fig:relationships_deps}.) Depending on the exact definition of the relations, the resulting representation may or may not exhibit the desired properties, or it may even exhibit properties not found in the actual system, but coming from externalities from the data, as discussed in Section~\ref{sec:data}.

For example, consider recording brain activity from an individual as they progress through different tasks (reading, watching a video, resting, etc.). Different tasks require the activation of distinct sets of brain regions. How do we define relations between brain regions? As depicted in Fig.~\ref{fig:relationships_deps}, we could define $k$ nodes to be related if a task requires all $k$ nodes to be active. Alternatively, we could define a relation between $k$ nodes if the $k$ nodes were found to co-activate during a task. Finally we could call two nodes related if they have a high enough measure of pairwise similarity, perhaps assessed by correlation or mutual information. Depending on our chosen definition of node relations, our resulting representation either will or will not encode a subset dependency. In this example, only the definition of node co-firing exhibits a subset dependency, which we could capture in a simplicial complex representation. Now consider a city bus system fragment including stations, roads, and bus lines (Fig.~\ref{fig:relationships_deps}, bottom). First, we could define a relation between $k$ nodes as the sets of stations along an entire bus route. That is, $k$ stations are related if they together form a whole bus route. This definition would propagate no subset or temporal dependencies to the representation. Second, we could instead call $k$ nodes related if they share at least one bus line. Consequently we now have a subset dependency that must be captured by our choice of representation. Third, we might define two bus stations as related if they are subsequent stops along a route. This third, inherently pairwise, definition of relation could be represented with a graph. Note that none of these three definitions encode the temporal dependency, which may or may not be present in the available data. For example, if we had access to, not only stations' locations, but also passenger trajectories within the system, we could encode the temporal dependencies using HONs.

\begin{figure}
    \centering
    \includegraphics[width =\textwidth]{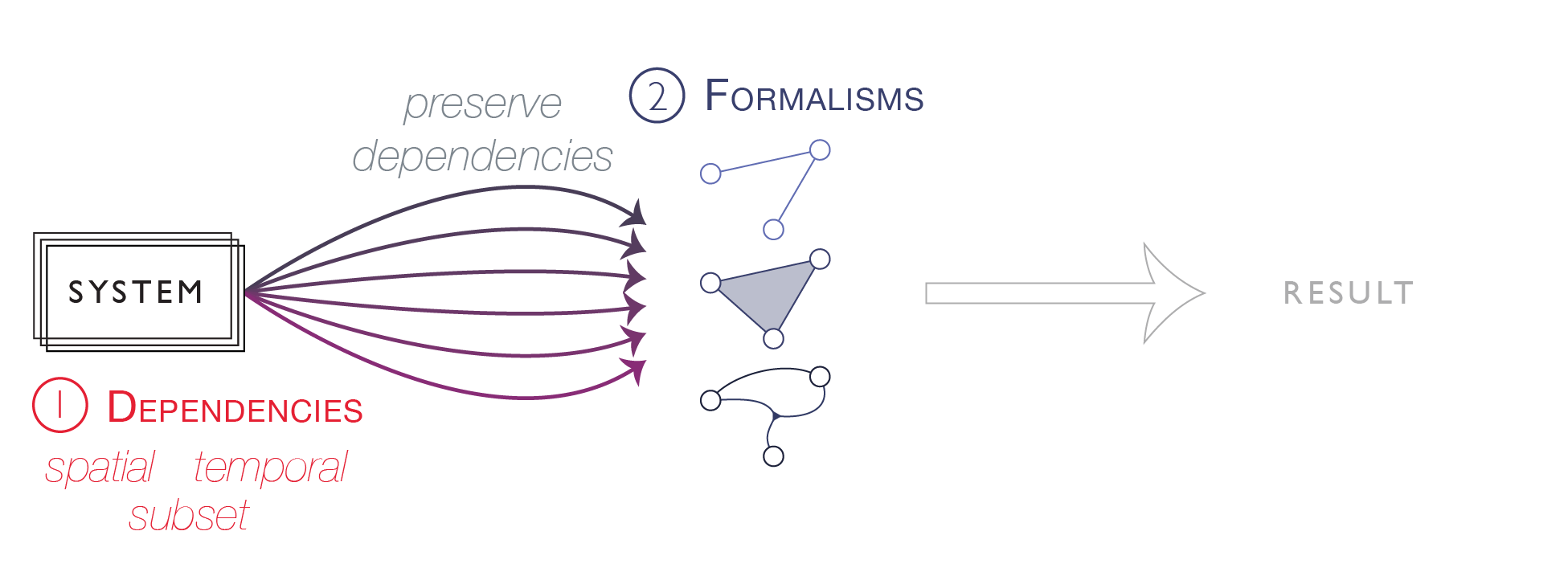}
    \caption{\textbf{Choosing a formalism marks the second step of our analysis pipeline.} Formalisms include graphs, simplicial complexes, and hypergraphs. We argue that the choice of formalism should be made in order to most faithfully capture system dependencies.}
    \label{fig:flow_forms}
\end{figure}

The above examples, and those in reference \cite{spivak2009higher}, illustrate the fact that one must carefully choose relations to effectively encode dependencies (Fig.~\ref{fig:flow_forms}), or, equivalently, that whether or not a given representation exhibits a dependency is a (sometimes subtle) question of semantics. This is to say, the modeling choices concerning relations, representations, and dependencies are highly, and unavoidably, interdependent on one another. We must be aware of what dependencies exist in the system, which of those are encoded or neglected in the representation, and which come from external sources. In the scientific community, these difficult choices are usually made following the common practices that we delineate next.

\subsubsection*{Encoding subset dependencies}
If a system exhibits subset dependency, it is common practice to use either simplicial complexes or hypergraphs to represent it. In the case when \emph{any} subset of a set of related units are also related, then an appropriate formalism is the simplicial complex, since this formalism has the downward inclusion property (see Section \ref{sec:group_dep}). In the terms used in Section~\ref{sec:group_dep}, the predicate $P$ is true for any subset of an existing relation. If instead only \emph{some} subsets of related units are related, then one could argue that a hypergraph is the appropriate formalism to use, since it allows for great freedom in encoding relations among subsets of related units. Equivalently, a particular subset dependency gives a particular choice of the predicate $P$, which in turn induces a particular hypergraph. Recall that the important difference between hypergraphs and simplicial complexes is the notion of a subedge. Drawing from Remark 3.5 of \cite{spivak2009higher}, if a 1-simplex $\{a,b\}$ and two 2-simplices $\{a,b,c\}$ and $\{a,b,d\}$ exist, then by definition $\{a,b\}$ is a sub-relation (formally called a \emph{face}) of both $\{a,b,c\}$ and $\{a,b,d\}$. However, if instead we had hyperedges $\{a,b\}$, $\{a,b,c\}$, and $\{a,b,d\}$ in a hypergraph, we cannot say if $\{a,b\}$ is a sub-relation (sub-edge) of $\{a,b,c\}$, $\{a,b,d\}$, both, or neither. This connection or lack thereof between relations and sub-relations crucially affects interpretation of the system representation.

\subsubsection*{Encoding temporal dependencies}
As discussed in Section \ref{sec:hons}, one way to encode temporal dependencies uses the idea of Higher Order Networks (HONs). Recalling our previous description, the HON begins with a set of paths, and from the patterns found therein creates a graph in which nodes correspond to ordered sets of units in the original system, and edges connect nodes based on temporal dependence. In this way, the HON takes the temporal dependency (for example, paths from A to B always lead to C), and encodes it in a special kind of node, derived from the original units of the system. At the time of writing, HONs have been defined for paths on graphs. It is still an open question how to extend the HON formalism to simplicial complexes or hypergraphs so that the resulting representation could exhibit both temporal dependencies and arbitrary subset (or spatial) dependencies.

\subsubsection*{Encoding spatial dependencies}
Possibly the most straight-forward method to encode spatial dependencies constructs a weighted graph in which the edge weights in some way represent how close or far nodes lie from each other. However, we highlight the fact that edge weights are a popular mechanism to also encode different kinds of information, and once we encode one piece of information within the edge weight we cannot then use edge weights to also encode spatial dependencies. For example, if we build a graph for the transportation system of a city, we may want to encode both traffic flow and road length. Usually, both types of information are encoded using edge weights, so we are left with three alternatives. The first is to choose an edge weight that aggregates both types of information. The second is to use a multilayer network (see \ref{sec:variations}) in which each layer has weighted edges reflecting a single type of relation \cite{chodrow2016demand}. The third is to create a more holistic representation that efficiently combines the spatial information, traffic flow, and road length while also including any interactions between edge types. This challenge is yet another example highlighting that data availability, system dependencies, and choice of representation are not independent of one another.

\paragraph*{} If the only challenge to the study of complex systems were the choice of representation, then our discussion would be near complete. However, real-world systems usually have at least two or more dependencies, including those we do not discuss in this paper. For example, the subway network (Fig.~\ref{sec:temporal_dep}) contains both temporal dependencies (evidenced in passengers' routes) and spatial dependencies (the routes taken are usually constrained by geographical proximity); while the coauthor system (discussed further in Section \ref{sec:examples}) could be further constrained by both temporal and subset dependencies. Moving forward, we will need to develop novel methods for systematically representing and encoding complex systems with multiple dependencies.


\section{Mathematical relationships between formalisms}

At this point it may seem that the choice of representation wholly restricts the perspective and possible analyses on the data. For example if we encode the data as a directed hypergraph, we can only perform analyses using hypergraph methods. However as each of these base formalisms record relations between nodes, perhaps we could utilize the underlying mathematical relationships between each of these formalisms to gain additional insights. In this section we will explore the formal mathematical relationships between graphs, simplicial complexes, and hypergraphs, and then we will discuss the assumptions needed or information lost as we move from one to another.

\subsubsection*{From hypergraph to simplicial complex: Forgetting independent sets}
\begin{figure}
    \centering
    \includegraphics{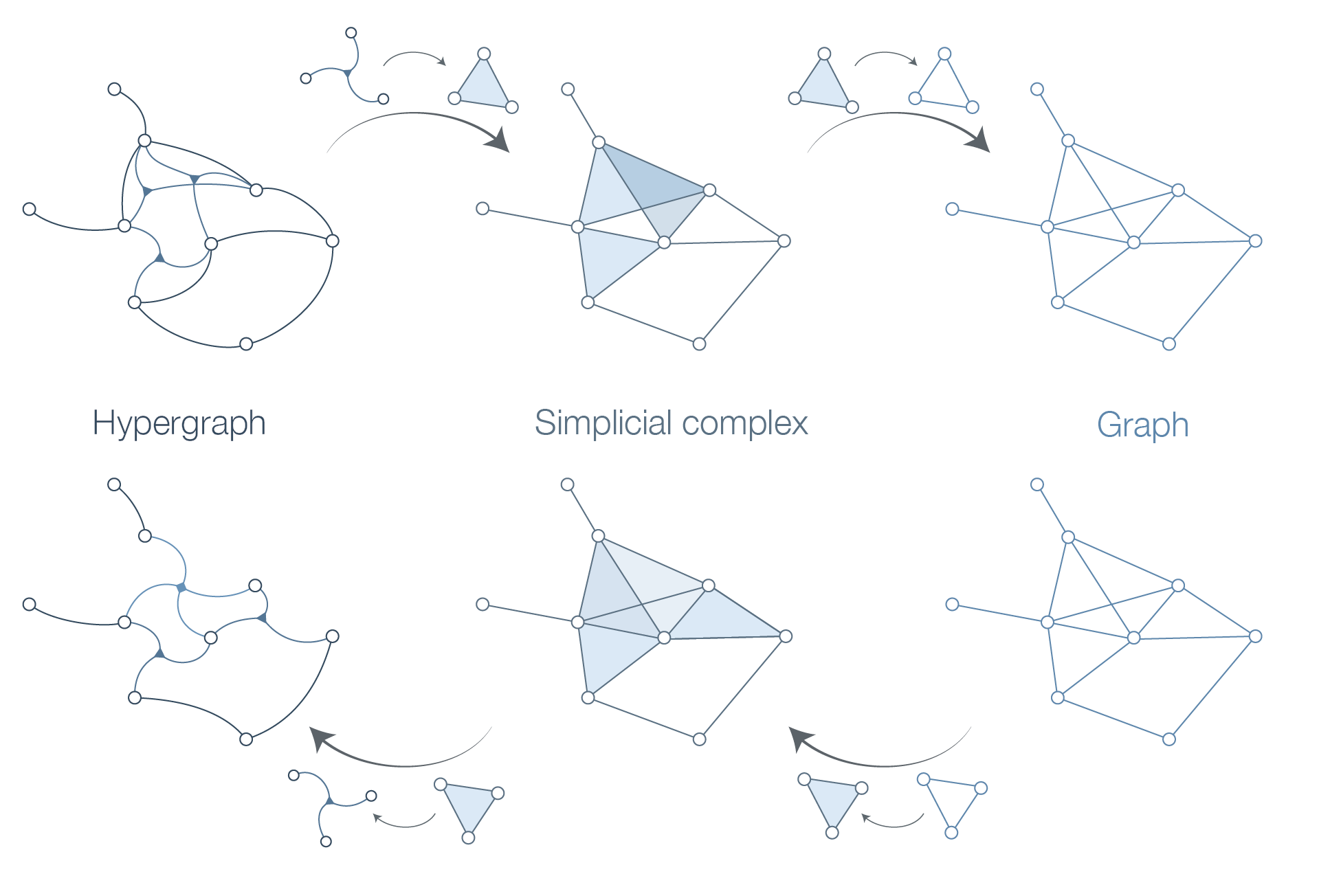}
    \caption{\textbf{Transitioning between formalisms requires added assumptions or engenders forgetting information.} \emph{(Top, left)} The original example system as a hypergraph. \emph{(Top, middle)} When we send hyperedges to simplices, we create a simplicial complex, and \emph{(Top, right)} by keeping all edges we form a graph. Going in the other direction, we begin with the same graph \emph{(bottom right)}, fill in all cliques as simplices to obtain a simplicial complex \emph{(bottom middle)}, and send maximal simplices to hyperedges to form a hypergraph \emph{(bottom left)}. Note that the hypergraphs on the top left and bottom left differ from one another.}
    \label{fig:relations}
\end{figure}

First let us imagine that from our data we have constructed a hypergraph $H$. If we would like to create a simplicial complex $K_H$ from $H$, we might first map the nodes of $H$ to nodes of $K_H$, before dealing with the hyperedges. Recall that in a simplicial complex we have simplices that connect multiple nodes, but we also have the downward closure restriction that if we have a simplex $r$, then any $r' \subseteq r$ must also be a simplex. So then to form $K_H$ we could take any hyperedge connecting $k+1$ nodes and form from it a $k$-simplex (Fig.~\ref{fig:relations} top left), thereby forcing the downward closure of the hyperedge relation so that the system representation can abide by simplicial complex rules. Additionally note that if we have a hyperedge $a$ on nodes $\{v_0,\dots, v_k\}$ as well as a hyperedge $b$ on a subset of these nodes, the simplicial complex will view $b$ as redundant information, since by definition every subset of nodes in $a$ will be connected by simplices. In this way, we say that the simplicial complex ``forgets'' the existence of $b$ as a relation observed independently of all other relations (specifically observed independently from the relation $a$). We can also see this forgetting notion in the matrix representation of the structure itself: from a hyperedge incidence matrix we only need to keep the maximal hyperedge rows in order to build the corresponding simplicial complex incidence matrix. Additionally, $K_H$ will also lose information regarding the total number of relations in which a node is involved, since many of those original hyperedge relations may be a subset of another hyperedge relation. On the other hand, this procedure allows us to access methods that are available for simplicial complexes but not for hypergraphs (discussed more in Section \ref{sec:methods}). Overall, in the hypergraph each hyperedge between a set of nodes arises independently, so that having additional hyperedges (or the lack thereof) between subsets of nodes within a larger hyperedge indeed supplies more information than the one largest hyperedge. In contrast, we can define a simplicial complex by its largest simplices (formally called maximal simplices) alone.

\subsubsection*{From simplicial complex to graph: Forgetting polyadic relations}
Next let us assume that we are given a simplicial complex $K$, and that from $K$ we wish to construct a graph $G_K$ that still represents our data. This transition is more straightforward, as we can take all of the 1-simplices of $K$ to be edges of the graph $G_K$. Said another way, if two nodes participate in the same $k$-simplex in $K$, then we draw an edge between these two nodes in $G_K$ (Fig.~\ref{fig:relations}, top right). By performing this transition from simplicial complex to graph, we are now forgetting polyadic relations between nodes. For example, in a simplicial complex we may have three nodes connected by three 1-simplices, or connected by three 1-simplices and a 2-simplex; in a graph, by contrast, we can only show these three nodes as being all-to-all connected by edges thus eliminating our ability to distinguish between the two cases. One can also move from a hypergraph to a graph by drawing an edge between two nodes only if the two nodes were connected by a hyperedge. The resulting graph recovered from this process will be the same as the graph obtained by moving from a hypergraph to a simplicial complex to a graph following the described protocol.

\subsubsection*{From graph to simplicial complex: Assuming polyadic relations}
What happens if we instead move in the other direction? What are the assumptions necessary to take a graph such as the graph shown in Fig.~\ref{fig:relations}, right, and construct from it a simplicial complex or a hypergraph? First, let us begin with a graph $G$ and construct a simplicial complex. If we make the assumption that all nodes involved in a $(k+1)$-clique of $G$ are related, then we can construct a simplicial complex $K_G$ by filling in each $(k+1)$-clique with a $k$-simplex. This particular construction is called the \emph{clique complex} \cite{kahle2009topology} or the \textit{flag complex} \cite{kahle2014sharp}, and is often denoted by $X(G)$ (Fig.~\ref{fig:relations}, bottom right). We reemphasize that for this construction, it is necessary to assume that all nodes within a clique are all together related as a single functional unit. Importantly this clique-to-simplex assumption may not be appropriate for all systems. One example arises from social conversations in which three people may converse only in pairs and never together as a three-person group.

\subsubsection*{From simplicial complex to hypergraph: Assuming that only maximal simplices are independent}
As we consider moving from simplicial complex $K$ to hypergraph $H_K$ we are faced with a few options. First, since a simplex by definition implies that all subsets of nodes within a simplex are also related, then we could take every simplex and form from it hyperedges between all subsets of nodes within the simplex. In constructing the hypergraph in this way, we carry through the downward closure restriction. Alternatively, we could perform a conversion more akin to the inverse of the hypergraph-to-simplicial-complex conversion discussed above by adding a hyperedge for each maximal simplex of $K$ (Fig.~\ref{fig:relations} bottom left). Assigning hyperedges only for maximal simplicies can be seen as a conservative approach; that is, we can uniquely define a simplicial complex using its maximal simplices so that in forming the new hypergraph, we are assuming the fewest number of hyperedges necessary to preserve only the polyadic relations with the most nodes.

\subsubsection*{From hypergraph to graph, and from graph to hypergraph}
Perhaps most importantly, note that from a graph we can move to a simplicial complex, then to a hypergraph, then back to a simplicial complex, and finally back to a graph following the translations discussed above. In this process, we will recover the original graph with which we began. However, the opposite is not the case. As depicted in Fig.~\ref{fig:relations}, we can begin with the hypergraph on the top left, move through the simplicial complex, to the graph, and then move back along the bottom row from right to left and we will in fact recover a very different hypergraph than the one from which we began. This exercise emphasizes the information lost or forgotten in moving down the formalism ladder. Specifically, since each hyperedge may arise independently of all others (most notably independently of any hyperedge that is a superset), we not only lose information when moving towards a graph but also cannot recover this information when moving back up from a graph to a hypergraph. 

\begin{figure}
    \centering
    \includegraphics[width = \textwidth]{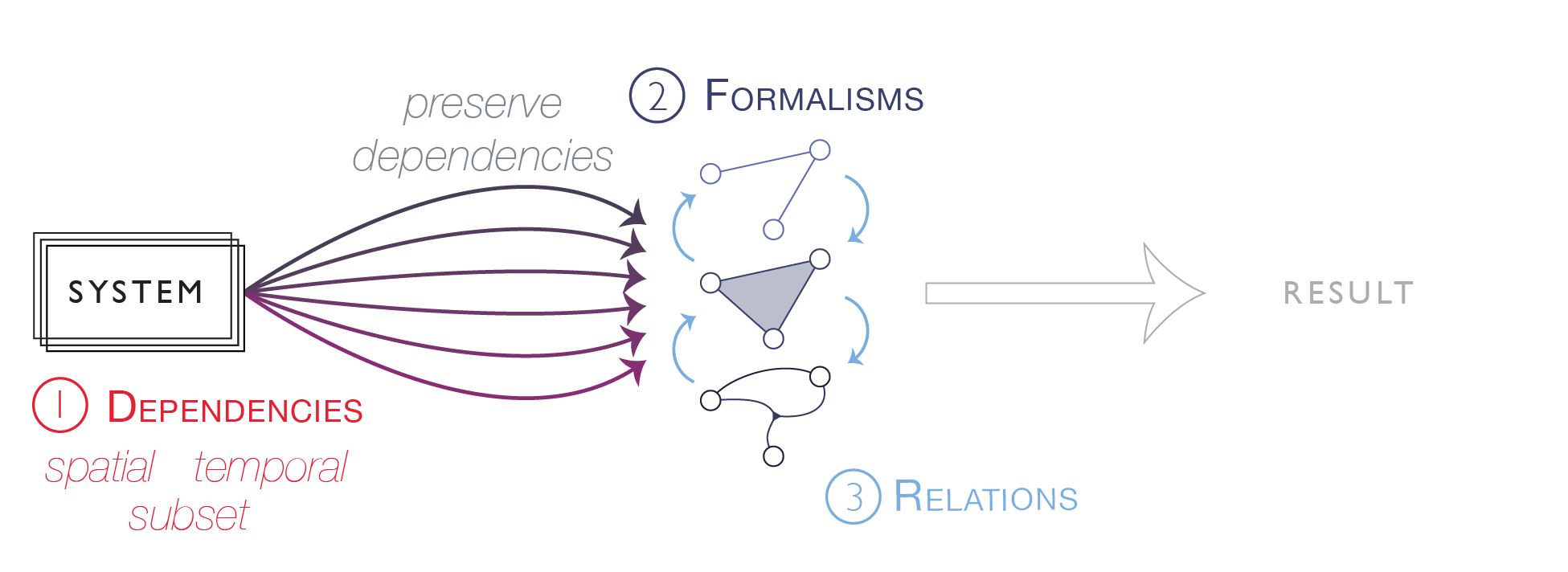}
    \caption{\textbf{Mathematical relations between formalisms add an optional step to the pipeline, only to be used with caution.} After an appropriate formalism has been chosen for system representation, one can make use of mathematical similarities between formalisms to view their representation from the lens of a different formalism. Importantly, we note that moving from hypergraphs to simplicial complexes to graphs can result in a loss of information, while the reverse direction can require one to make assumptions about the system.}
    \label{fig:org_relations}
\end{figure}

\paragraph*{}
We note that the above protocols of moving from one formalism to another do not encompass all possibilities. One could define a simplicial complex from a graph by simply keeping all edges as the 1-skeleton and having no larger simplices. Or perhaps one might form a weighted graph from a hypergraph by assigning edge weights as some function of the hypergraph structure \cite{neubauer2009towards,chodrow2020annotated,habibi2018disruption}. Though we here discussed moving between formalisms as the third step in the pipeline (Fig.~\ref{fig:org_relations}), moving from one formalism to another \emph{after} the initial encoding of data into a formal representation should be performed only with extreme care, as any translation requires adding assumptions or forgetting relations or independencies.

\section{Methods suitable for each representation} \label{sec:methods}


Now that we have exerted the effort necessary to properly represent our data as a graph, simplicial complex, or hypergraph, how do we analyze the resulting structure? In this section, we will describe methods that can be used to evaluate precisely how each of the three base formalisms offer unique perspectives on the system under study. We recognize that many such methods exist, but for clarity we will focus on a few techniques that help us identify similarities among and differences between representations.

Before we begin, we briefly provide another a note of caution. The fact that a method of interest might currently intake only one particular formalism does not justify the use of that formalism in representing our data. To further illustrate the point, if we intend to understand the spread of a disease by way of calculating the epidemic threshold \cite{ChakrabartiWWLF08}, we would find that existing methods to calculate the epidemic threshold do so from a graph representation. The theory of disease spread on hypergraphs and simplicial complexes currently is a nascent area of research \cite{iacopini2019simplicial,jhun2019simplicial}, so one might not find a definition of the epidemic threshold that uses either of these polyadic formalisms in the literature and is appropriate for the system at hand. Nevertheless, the absence of this particular notion for polyadic formalisms does not imply that we are justified in using a graph formalism to represent the system. Generally, a result is unlikely to offer fruitful insight into a system if the calculation was performed on a representation that itself is ill-suited for the system.

\begin{figure}
    \centering
    \includegraphics{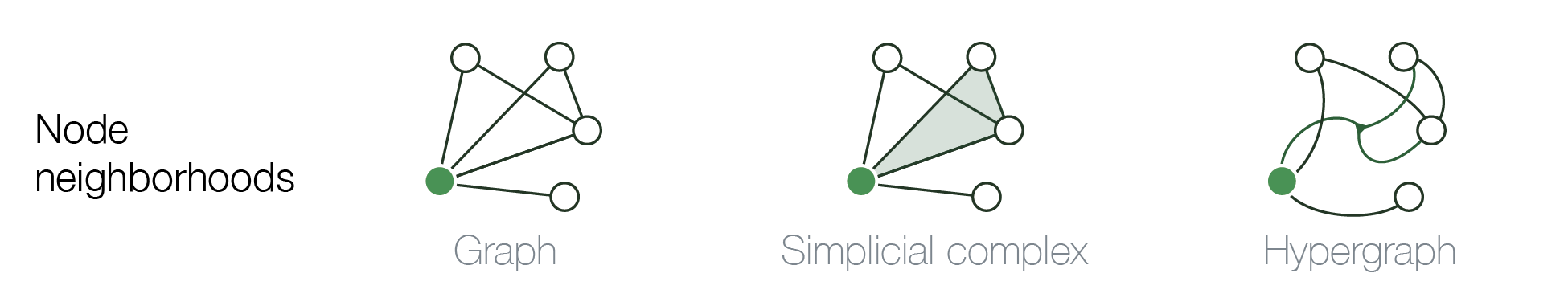}
    \caption{\textbf{Formalisms provide different perspectives on the neighborhood of a node.} \emph{(Left)} The colored node has four direct neighbors and participates in two triangles. \emph{(Middle)} The colored node has four direct neighbors and participates in one 2-simplex. \emph{(Right)} The colored node has two neighbors connected to itself exclusively, and two neighbors accessible through a larger hyperedge.}
    \label{fig:table1}
\end{figure}

\subsection{Methods for graphs}\label{sec:methods-for-graphs}

As the most well-known of the three formalisms in data analysis, graphs have offered scientists interpretable and easily-computable tools for centuries. Thanks to this rich history of graph analysis, we can computationally investigate graphs at many different levels: the local node or node-neighborhood level, a meso-scale level to see larger patterns, and the global level to summarize the entire object. Though myriad metrics exist, for the sake of brevity we limit our discussion below and point the interested reader to \cite{newman2018networks} to learn more.

Analysis methods on graphs are relatively well-developed and expansive, reflecting their long and wide-spread employment in complex system research. In using them, we can learn a great deal about the system's organizational structure by understanding the neighborhood of nodes within the graph (Fig.~\ref{fig:table1}, left). At the most basic level, the number of edges incident to a node $v_i$ is called the node \emph{degree} and is denoted $k_i$. The distribution of degrees can constrain the graph's large-scale organization, for example tracking the emergence of a giant connected component \cite{molloy1995critical}. At the neighborhood level, we can investigate measures of the connectivity between a node's neighbors. A common example is the \emph{clustering coefficient} $c_i$ of a node $v_i$. Formally the clustering coefficient is 
\begin{equation}
    c_i = \frac{2\mu_i}{k_i(k_i-1)} \,,
\end{equation}

\noindent where $\mu_i$ is the number of edges between neighbors of $v_i$. The numerator counts the number of triangles in which $v_i$ participates and the denominator normalizes by the number of triangles that could possibly form around $v_i$. Broadly, the degree and clustering coefficient are examples of a much broader class of metrics proposed for the description of local and neighborhood structure in graphs.

\begin{figure}
    \centering
    \includegraphics[width = \textwidth]{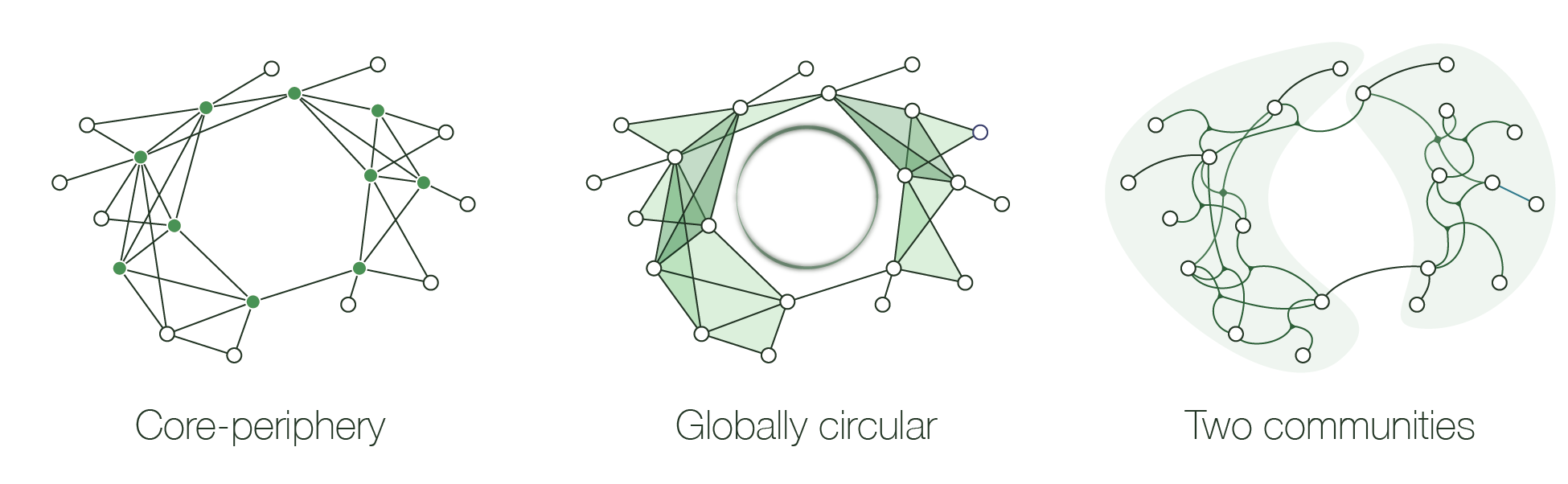}
    \caption{\textbf{The three formalisms and their corresponding downstream analyses can offer different perspectives on a complex system.} \emph{(Left)} For this example system, a graph representation suggests a global core-periphery organization. \emph{(Middle)} A simplicial complex representation of the same system appears to show a globally circular structure. \emph{(Right)} The hypergraph representation of the same system hints at the presence of two communities.}
    \label{fig:methods2}
\end{figure}

Complementing such descriptions, other statistics have been defined to measure the nature of paths in the graph and markers of meso-scale structure. For example, the average path length, various types of centrality \cite{freeman1977set,bavelas1950communication}, notions of modularity \cite{guimera2004modularity,krzakala2013spectral,palla2005uncovering}, and the property of small-worldness have proven useful in the study of a wide variety of systems from the human brain \cite{bullmore2009complex,bassett2017network} to granular materials \cite{papadopoulos2018particles}. One particular statistic that, at the time of writing, we found to be unique to the graph formalism, is a measure of core-periphery structure. A graph with core-periphery structure contains a dense group of nodes connected to each other called the \emph{core}, and a second group of nodes called the \emph{periphery} that mostly connect to the core rather than to other nodes in the periphery \cite{borgatti2000models,rombach2014core} (Fig.~\ref{fig:methods2}, left). For a description of other network measures, we refer the interested reader to prior literature \cite{newman2018networks,rubinov2010complex}. Additionally, we note that in real world systems, the values of many of these network statistics are statistically correlated with one another over instances in a graph ensemble, and these patterns of shared variance can be used to distinguish between types of systems \cite{costa2007characterization,onnela2012taxonomies}.

\subsection{Methods for simplicial complexes}

 Simplicial complexes entered the data analysis scene more recently than graphs. Yet, we can still use a simplicial complex to investigate multiple levels of system architecture with intuitive measures. As with graphs, we keep this section brief by focusing only on a few basic measures and then one measure that is unique to simplicial complexes.

We might first seek to extrapolate basic graph definitions to simplicial complexes. If we view a graph as the 1-skeleton of a simplicial complex, then the graph degree of node $v_i$ is the number of 1-simplices in which $v_i$ participates. By extending this idea, we can understand the neighborhood of a node (Fig.~\ref{fig:table1}, middle) by defining the \emph{simplex participation} of node $v_i$ as the vector $P(v_i)$ in which the $k^{th}$ element is the number of $(k-1)$-simplices in which $v_i$ participates. One could also record the vector of simplices in which the node participates (called the upper degree in \cite{serrano2020simplicial}), or the number of simplicies not contained in any larger simplices, i.e. the maximal simplices, in which the node participates \cite{sizemore2018cliques}. Similarly, we might ask whether and how the clustering coefficient could be extended to the simplicial complex formalism. Depending on the precise properties that one intends to capture, one could use a ratio of simplices from dimensions $k$ and $(k-1)$ to formalize the notion of a clustering coefficient. However, in the simplicial complex formalism each simplex can be considered as a fundamental building block, so it makes sense to also define a clustering coefficient for an arbitrary $k$-simplex as in \cite{maletic2008simplicial}. In a complementary effort, the notion of centrality has recently been extended from the graph formalism to the simplicial complex formalism \cite{estrada2018centralities}.

In addition to extending graph measures to simplicial complexes, we can also harness underlying algebraic topology to uncover more complicated motifs within the system. The downward closure requirement within the simplicial complex definition gives us the ability to accurately identify which simplices are involved in higher dimensional simplices. Consequently we can then detect where a dearth of simplices leaves topological voids in the complex (Fig.~\ref{fig:methods2}, middle). Detecting topological voids is the work of \emph{homology}, and as homology relies on well-defined mappings from larger to smaller simplices, this method is best suited for the formalism of simplicial complexes.\footnote{Sometimes the words ``structural'' and ``topological'' are used interchangeably. In this work, we use the adjective ``topological'' to modify nouns relating to the theory of algebraic topology. We use ``structural'' to generally refer to the patterns formed by the units and relations of a system.}

\begin{figure}
    \centering
    \includegraphics{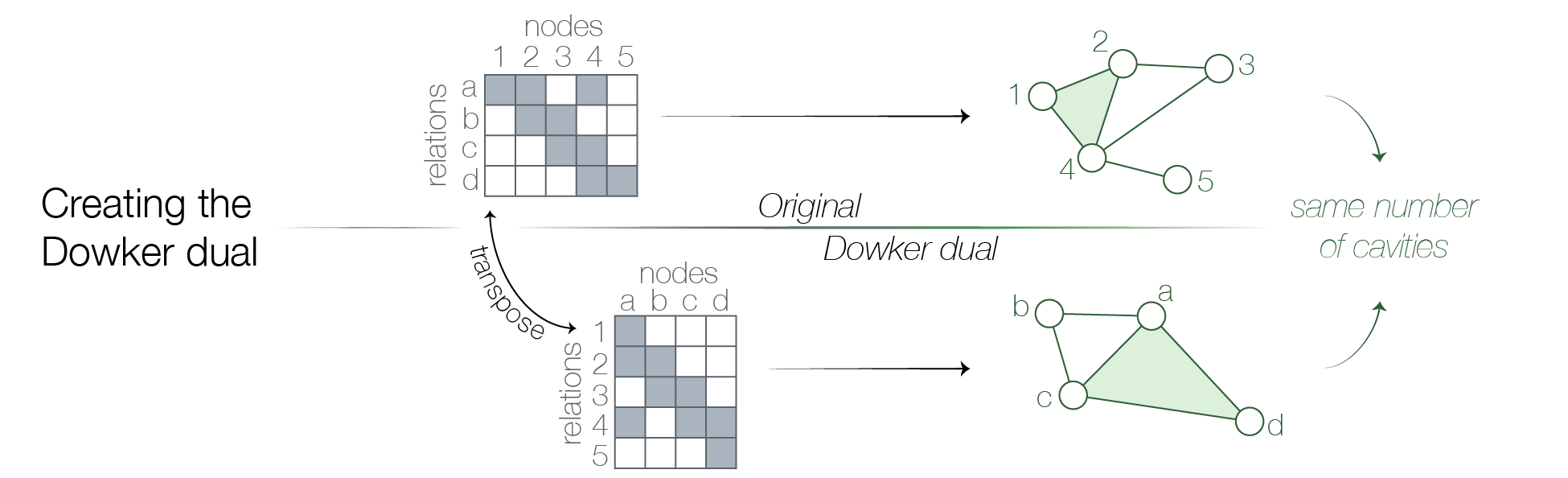}
    \caption{\textbf{Constructing the Dowker dual of a simplicial complex.} Given a concurrence matrix denoting which nodes connect via relations (top left), we can create a simplicial complex with each simplex defined by a relation ($a,b,c,d$). Alternatively, we could transpose the matrix so that now we have four nodes ($a,b,c,d$) and five relations (bottom left). From this transposed concurrence matrix we can then create a simplicial complex whose simplices are defined by relations in the new concurrence matrix (bottom right). This simplicial complex is called the Dowker dual of the original simplicial complex, and will have the same number of topological cavities.}
    \label{fig:dowker}
\end{figure}

Simplicial complexes can also be ``reversed'' in a way that can be useful in understanding the structure of grouped nodes, while preserving the topological organization of the system. Consider constructing, for example, a simplicial complex in which nodes represent neurons of the zebrafish brain, and simplices represent co-activity during a task. We could encode the complex as a $\#simplices \times \#nodes$ binary matrix sometimes also called a concurrence matrix \cite{giusti2016two,dowker1952homology}. In the top left of Fig.~\ref{fig:dowker} we show a small example concurrence matrix of five nodes ($1,2,3,4,$ and $5$) connected through four possible relations ($a,b,c,$ or $d$). We create a simplicial complex (Fig.~\ref{fig:dowker}, top right) by drawing maximal simplices between nodes that share a relation in the concurrence matrix. For the zebrafish example, the simplicial complex could contain relatively few simplices but orders of magnitude more nodes (depending on data availability of course), making calculations cumbersome. As an alternative, we could ``reverse'' the structure by constructing the \textit{Dowker dual} \cite{dowker1952homology}. Here, the role of nodes is swapped with the role of relations \cite{dowker1952homology}. In the zebrafish example, we would form a node for each co-activity relation, and then connect two nodes by simplices if they share a participating region of the zebrafish brain. In Fig.~\ref{fig:dowker} we transpose the concurrence matrix to swap the role of nodes and relations, and then show how we again create a simplicial complex now called the Dowker dual. This new complex will have the same number of nodes as the original complex had maximal simplices, so that if the number of relations was small with respect to the number of nodes in the original complex, studying the Dowker dual will be more computationally tractable.\footnote{This construction is akin to the \textit{line graph} construction in graph theory \cite{harary1960some}.} Importantly, studying the Dowker dual preserves specific topological structure within the system \cite{dowker1952homology}: the homology groups of a simplicial complex and its Dowker dual are isomorphic. Thus, the Dowker dual can be an incredibly efficient representation, assuming that we still respect the scientific question at hand. 

\subsection{Methods for hypergraphs}\label{sec:methods-for-hypergraphs}

Like simplicial complexes, hypergraphs have gained popularity only recently, as the field has begun to realize the importance of encoding polyadic relations in systems. The hypergraph formalism is a natural extension of the graph formalism, so unsurprisingly many (though certainly not all) computational methods for hypergraphs are extensions of computational methods for graphs. As with the other formalisms, we will highlight basic methods here as well as a method unique to hypergraphs.


The formalism of hypergraphs is also complemented with a set of descriptive statistics. Importantly, recall that each hyperedge arises independently since we have no rules relating hyperedges to each other, and consequently computations on hypergraphs must be interpreted differently from related computations on simplicial complexes. Starting simply, we can first extend the concept of degree to hypergraphs. In a hypergraph, the degree of a vertex $d_H(v_i)$ is the number of hyperedges containing $v_i$, sometimes called the \emph{hyperdegree}. Since hyperedges can connect any number of nodes, we also define the \emph{hyperedge cardinality}, also called the \emph{hyperedge degree}, as the number of nodes contained by the hyperedge. Importantly, note that in the definition of vertex degree, we do not stratify by hyperedge cardinality as the appearance of a large hyperedge gives no information about the existence of smaller hyperedges. Instead, a large hyperedge is simply another relation that contains our node of interest.

In order to understand a node's neighborhood in a hypergraph (Fig.~\ref{fig:table1}, right), next we move to a definition of the hypergraph clustering coefficient (see \cite{klamt2009hypergraphs,estrada2006subgraph,pena2012bipartite,gallagher2013clustering} for others). Recall the graph clustering coefficient measures connectivity of a node's neighbors via connections that \emph{do not} include the node of interest. If we examine, for example, node $v_i$ and its neighbors, some neighbors will be connected via hyperedges that do or do not include $v_i$. Intuitively, node $v_i$ should have high clustering if its neighbors connect via hyperedges that do not contain $v_i$. The \emph{extra overlap} $EO(v_i)$ of a node $v_i$ helps us to quantify this idea; formally, the extra overlap of two hyperedges $e_j, e_k$ is defined as 
\begin{equation}
    EO(e_j,e_k) = \frac{|N(D_{j,k}) \cap D_{k,j}| + |D_{j,k} \cap N(D_{k,j})|}{|D_{j,k}|+|D_{k,j}|},
\end{equation}

\noindent where $D_{j,k} = e_j - e_k$ and $N(U)$ is the set of all nodes that are neighbors of any node within the set $U$. Then intuitively the extra overlap between two hyperedges counts the number of nodes connected by outside hyperedges, and we normalize by the size of the two hyperedges under consideration. Note that if we have only hyperedges of cardinality 2, then the extra overlap over two edges involved in a triangle is 1. Finally, the hypergraph \emph{clustering coefficient} $C_H(v_i)$ of a node $v_i$ is

\begin{equation*}
C_H(v_i) = 
\begin{cases}
  \dbinom{|M(v_i)|}{2}^{-1}\sum\limits_{e_j,e_k \in M(v_i)}EO(e_j,e_k) \text{  if $d_H(v_i) > 1$}\\      
  0 \text{  if $d_H(v_i) = 1$}
\end{cases}
\end{equation*}

\noindent where $M(v_i)$ is the collection of hyperedges that include $v_i$ \cite{zhou2011properties}. This definition for the hypergraph formalism is thus similar in spirit to the definition of a clustering coefficient for a graph. Indeed, the former is equivalent to the latter when all hyperedges have cardinality 2.

We note that the hypergraph also has the ability to uniquely represent the absent substructures of a system. Much like identifying repeated structural patterns (or \emph{motifs}) in a graph, a hypergraph allows us in principle to identify repeated patterns of \emph{absent} hyperedges. We may have a case where, for example, pairwise hyperedges exist between four nodes that also connect via a 4-hyperedge, but no 3-hyperedges exist. An interesting research question for such a representation is to ask why we observe a lack of three-node relations but an abundance of 2-node relations within every 4-node relation. Note that neither graphs nor simplicial complexes allow for this line of questioning due to the lack of polyadic relations or the requirement of downward inclusion, respectively. A detailed investigation of these absent substructures is outside of the scope of this paper; yet, we can take a step in that direction by defining the following statistic, which we call the \emph{fill coefficient} of a hyperedge $h$, as

\begin{equation}
f(h) = \frac{|g \in E: g \subsetneq h \text{ and } |g| > 1|}{2^{|h|} - 2 - |h|},
\end{equation}

\noindent where $E$ is the set of hyperedges, and $|\cdot|$ is the cardinality of a hyperedge. The fill coefficient intuitively describes the fraction of smaller hyperedges that exist within hyperedge $h$, taking into account the hyperedge cardinalities.

\subsection{Methods and dependencies}

Before closing this section, we note that both in choosing analyses and in interpreting results, we need to keep in mind the dependencies within the system. For example, after creating a simplicial complex from our data, how do we interpret its clustering coefficient? Or what does the diameter of a system mean when we have hyperedges of different cardinalities linking nodes instead of (dyadic) edges? How do communities found from a simplicial complex \cite{billings2019simplex2vec} with subset dependencies differ from communities found within a hypergraph \cite{kim2017community} without such dependencies? Can we intertwine different sorts of system dependencies to understand their impact on function? Examples of such intertwining methods include (i) Rentian scaling, which formalizes the interaction between structure and geography \cite{christie2000interpretation,bassett2010efficient,papadopoulos2018comparing}, and (ii) modularity maximization with spatial null models \cite{expert2011uncovering,betzel2017modular}. Careful consideration of the above questions can only lead to better motivated, more interpretable, and insightful results.

\begin{figure}
    \centering
    \includegraphics[width = \textwidth]{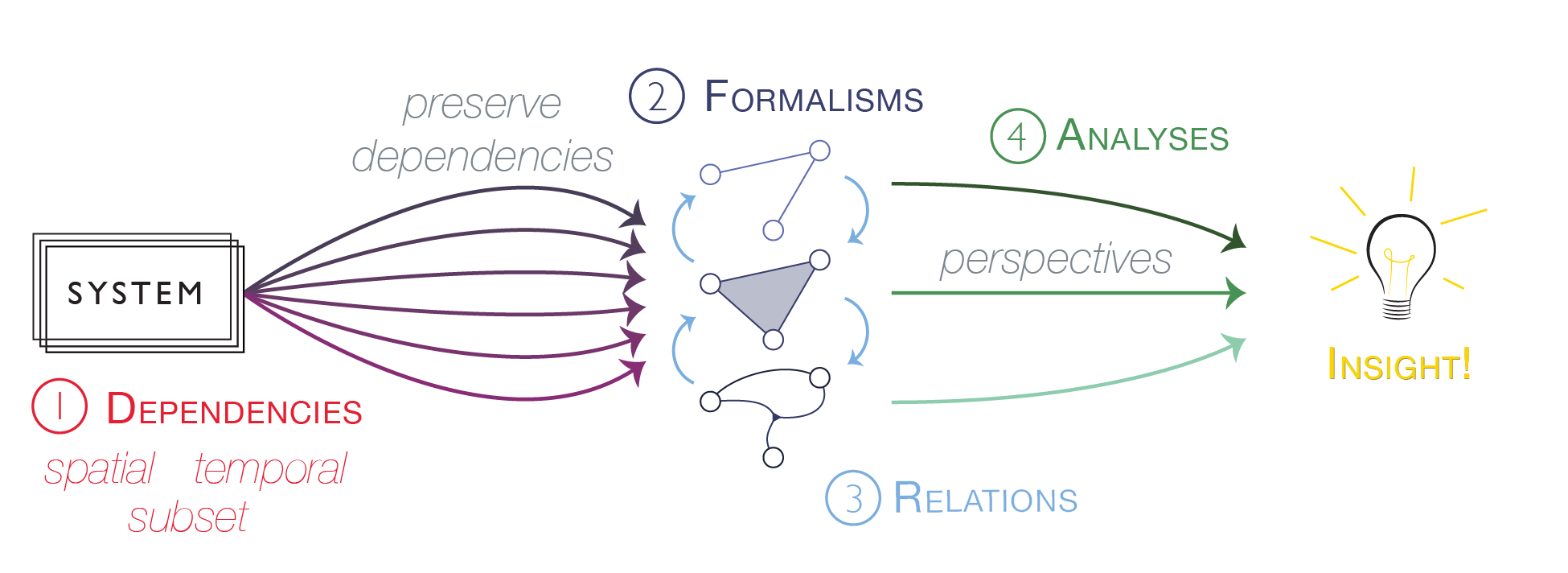}
    \caption{\textbf{Updated analysis pipeline includes consideration of which computational methods to perform on the chosen representation, and what distinct or complementary perspectives these methods offer.} The last step in our pipeline involves computationally analyzing the system representation. We note that each analysis provides its own perspective on the system representation. We recommend performing steps 1-4 with careful consideration in order to gain real insight into the system.}
    \label{fig:org_methods}
\end{figure}

To summarize, we see that each base formalism offers a particular perspective on the data it encodes. As we show in Fig.~\ref{fig:methods2}, the choice of formalism can influence how we interpret the complex system structure. The graph representation could suggest a core-periphery structure; the simplicial complex representation lets us see that globally the system organizes around one circle; and the hypergraph representation highlights the existence of two communities. We close this section by emphasizing the importance of formalism choice for proper representation of the data, and the appreciation that each analysis performed or pipeline chosen offers a different perspective on the underlying complex system (Figure~\ref{fig:org_methods}).

\section{Examples}\label{sec:examples}

Putting it all together, in this section we will discuss examples of a system, its dependencies, and how we might represent the system using the formalisms described above. We will then explore possible analyses on each representation and compare results. Importantly, we will see that the subtle differences in representations and their definitions can lead to inconsistent results and conflicting interpretations.

\subsection{Coauthorship}
\begin{figure}
    \centering
    \includegraphics{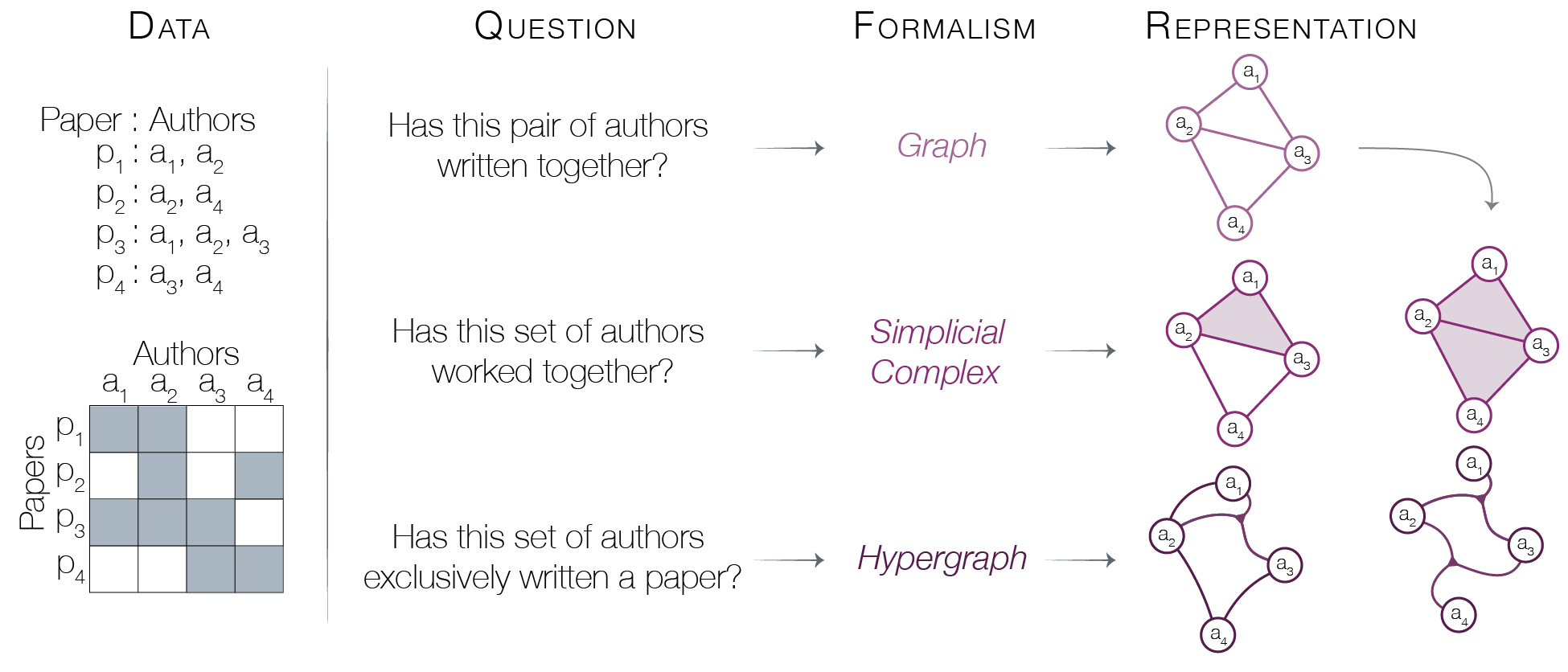}
    \caption{Example of different perspectives offered by each formalism on a co-author dataset. The data (far left) consists of a list of papers and their author list. Based on a question about what defines relations between authors, we build either a graph, simplicial complex, or hypergraph (right). If we started with the graph, we could also use the relations between formalisms to create a simplicial complex or hypergraph (far right), though this process can result in inaccurate representations of the original data.}
    \label{fig:example1}
\end{figure}

\begin{figure}
    \centering
    \includegraphics{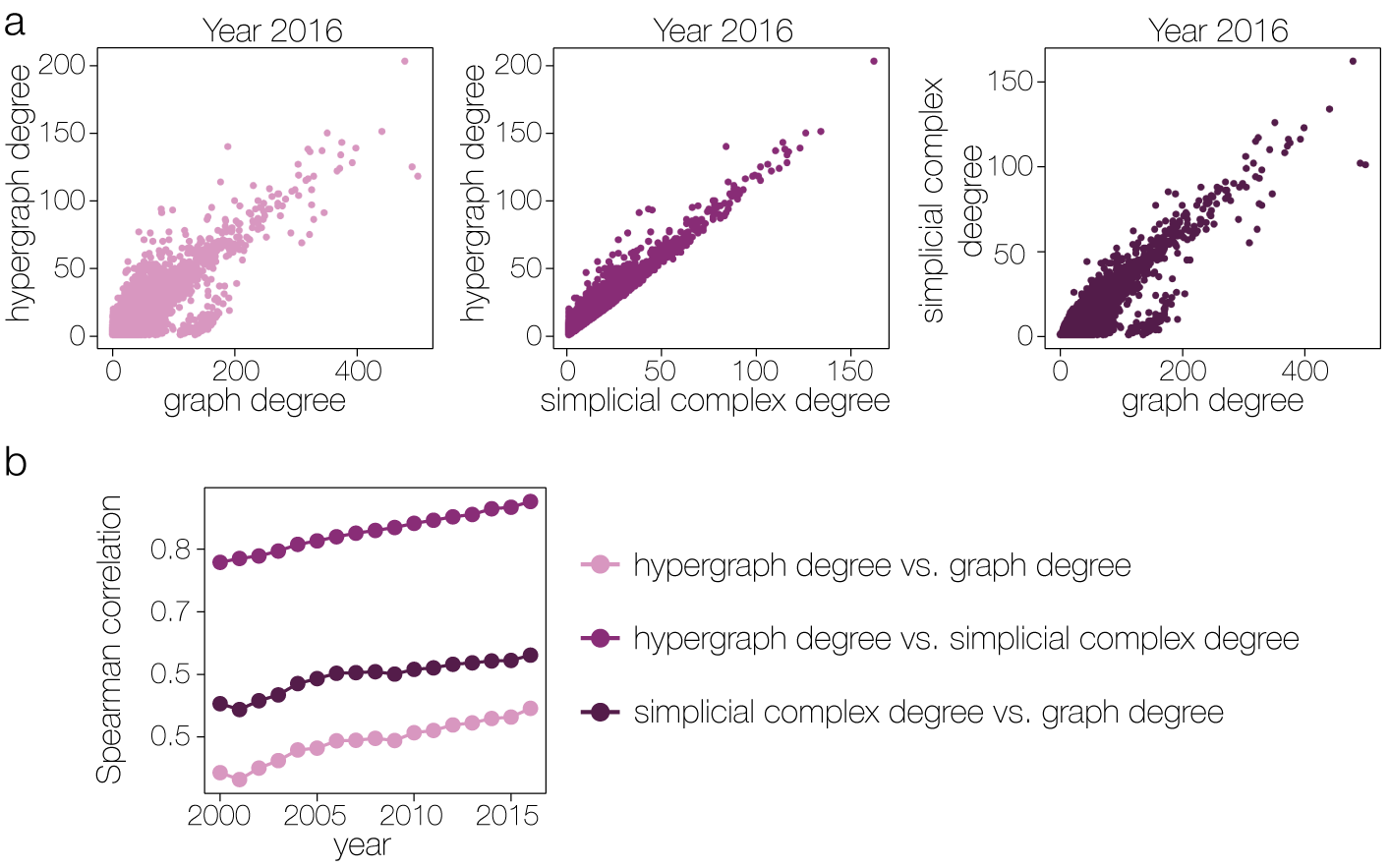}
    \caption{Correlation among degree measurements in different representations of the same dataset. \emph{(a)} Comparing the degree calculated from the graph or hypergraph representation (left), from the simplicial complex or hypergraph representation (middle), and from the graph or simplicial complex representation (right) from the coauthorship dataset extracted from the DBLP computer science bibliography database in year 2016. Correlation is relatively high for nodes of large degree, but relatively low for nodes of small degree. \emph{b} Spearman correlation coefficient calculated between node degrees from pairs of data representations in each year.}
    \label{fig:results1}
\end{figure}

Consider the system made up of scientific researchers who interact to write scientific papers (for example, \cite{clauset2017data}). What kind of dependencies govern the relations in this system? How should we encode this system formally? In the following paragraphs we analyze a fragment of this system following the workflow of Fig.~\ref{fig:org_methods}. We begin with a toy example (Fig.~\ref{fig:example1}) and later perform similar analyses on a real dataset (Fig.~\ref{fig:results1})

\paragraph*{Dependencies} First, we may expect to find spatial dependencies in this system, as the country of origin or university affiliation of a researcher may dictate which of their colleagues are willing or able to collaborate. Second, we may also expect temporal dependencies, as a researcher's past collaborators may also influence any future collaborations. Third, whether this system exhibits subset dependencies depends upon which precise fragment we are interested in studying. If we focus solely on authors and consider two or more authors as related whenever they have worked together at some point, then it is necessarily the case that whenever a group of three authors have coauthored a paper, then any two of them have coauthored a paper, and therefore a polyadic relation always implies all dyadic sub-relations (and all smaller polyadic relations). On the other hand, we may choose to focus on both researchers and scientific papers, in which case, we may want to think about one scientific paper as determining a single relation. In this scenario, the fact that three researchers are involved in a relation (because they have authored a paper all together) does not imply that two of them have authored (another, separate) paper together. We will keep these dependencies in mind as we move through the later analysis steps.

\paragraph*{Externalities: data availability}
In a vacuum, we may expect to see all of the aforementioned dependencies in this system. However, the data available may be biased in such a way that, for example, researchers working (and papers produced) in one particular country are over-represented. In this case, the data available may not adequately record the spatial dependencies involving, for example, researchers who travel frequently between two different countries. Moreover, if the dataset is further biased to include solely researchers that work in one particular institution, then it is possible that no spatial dependencies are recorded at all, since researchers in one institution may all work with one another with the same likelihood; that is, the location of one existing collaboration offers no new information about the likelihood of another collaboration. Importantly, access to the full data about researchers and scientific papers would allow us to build naturally any of the three formalisms discussed. However, if we only have information about co-authorship (which sets authors have worked with each other) rather than full knowledge of the data, we would only have been able to construct the graph version or the simplicial complex version, but not the hypergraph version\footnote{One could build a hypergraph version, but it would not be an appropriate representation for this scenario because it does not respect the blatant subset dependency.}.

\paragraph*{Externalities: research question}
Let us introduce a toy example to accompany our discussion. Consider a coauthorship dataset including four authors $a_1,a_2,a_3,$ and $a_4$ who have written four papers $p_1,p_2,p_3,$ and $p_4$ (Fig~\ref{fig:example1}, top). The three papers were authored as follows: paper $p_1$ was authored by $\{a_1,a_2\}$, paper $p_2$ by $\{a_2,a_4\}$, paper $p_3$ by $\{a_1,a_2, a_3\}$, and paper $p_4$ by $\{a_3,a_4\}$. This toy example illustrates that whether the chosen representation reflects the dependencies inherent to the system is sometimes a subtle question of semantics. For example, if the relations in our representation are defined using the first question (``has this pair of authors worked together on at least one paper?''), then the set of relations will necessarily exhibit a subset dependency, regardless of whether or not the data available records a subset dependency found in the real system. Said another way, one must be aware of which dependencies reflected in our representations come from the system, from how the data was collected, or from the representation constructed. In this example it is the question at hand, the intricacies of the system under study, and the data available that all together guide the choice of representation of these data. 

\paragraph*{Representations}

If we take this information and construct the classic coauthor network in which an edge exists between two authors if they have appeared as coauthors on a paper, then we recover the graph shown in Fig.~\ref{fig:example1}, top right. In particular, note that as we construct the co-authorship graph, we ask the following question exactly once for each potential relation: ``Has this \emph{pair} of authors worked together on at least one paper''? Alternatively, we can consider polyadic relations between authors and ask ``has this \emph{set} of authors worked together on a paper?'' This question naturally yields a simplicial complex (Fig.~\ref{fig:example1}, middle right), in which nodes form a simplex if the corresponding authors are a subset of the authors of at least one paper. Finally, we imagine the author list of the paper is non-redundant so that one paper corresponds to exactly one relation. Said another way, we respect that without each and every author, the paper could not have been completed. If we take this point of view, we will instead construct a hypergraph by repeatedly asking ``Has this set of authors exclusively (needing no other authors) written a paper together?'' This approach retains the large group of three authors, but now clearly shows that, for example, authors $a_1$ and $a_3$ have not worked on a project as an exclusive group. Note that this information is not recoverable from either of the other representations.

\paragraph*{Methods and Analyses} Next we analyze the three different system representations. Of our coauthor representation (graph, simplicial complex, or hypergraph), we might first ask a simple question about the involvement of a node (author) in paper writing in order to gauge the author's productivity. In the graph, we might use the node degree to recover this information, which would tell us that authors $a_2$ and $a_3$ participate in the same number of collaborations. Moving to the simplicial complex, we see by looking at node participation in maximal simplices that again authors $a_2$ and $a_3$ could be described as equivalently collaborative. However, in the hypergraph representation, if we look at node degree we see clearly that $a_2$ has participated in more collaborative projects than any other author, a conclusion that we are only able to draw from the hypergraph representation. A similar experiment comparing authors $a_1$ and $a_4$ shows that in this scenario, both the simplicial complex and hypergraph encodings view these authors as having different sizes of collaborative projects, while the graph structure does not. Specifically, the graph tells us that both $a_1$ and $a_4$ have worked with $a_2$ and $a_3$; the simplicial complex tells us that $a_1$ worked collectively with $a_2$ and $a_3$ whereas $a_4$ only worked individually with $a_2$ and $a_3$; and finally the hypergraph tells us that $a_1$ had an individual project with $a_2$ as well as a team project that also included $a_3$ while $a_4$ only worked on two-person papers. These analyses illustrate how the subtle differences in the three discussed representations and associated downstream analyses can yield insights that may be at odds with one another.

\paragraph*{Relationships between formalisms}
In this toy example we constructed each of the three representations directly from the data itself, with full knowledge of the raw data. However, we could also imagine that we are given the data already represented as one formalism and then try transforming our representation to another formalism. If we begin with one representation and translate to another formalism, will we recover the same information as if we had constructed the structure directly from the data? Here if we begin from a graph and move to a simplicial complex by attaching simplices to cliques (i.e. construct a \textit{clique complex}), we recover a simplicial complex with two maximal simplices formed by $a_1,a_2,a_3$ and $a_2,a_3,a_4$ (Fig.~\ref{fig:example1}, far right). Moving then from simplicial complex to hypergraph we would form a hypergraph with two hyperedges between $a_1,a_2, a_3$ and $a_2,a_3,a_4$; see Fig.~\ref{fig:example1}, far bottom right. If we asked the same questions about author participation as we did above, then we would find that both pairs ($a_2,a_3$ and $a_1,a_4$) now seem to contribute in exactly the same way across representations. In studying complex systems we may receive only one representation of the system rather than the raw data, which can make switching to a different representation that perhaps better suits the planned analyses enticing. However, the present exercise underscores the importance of understanding the assumptions made by each formalism; care must be taken when moving between formalisms, not simply in recasting the mathematical language used, but also in remaining true to the original data.

\paragraph*{Real dataset example}
We close this example by illustrating the above points in a real coauthorship dataset extracted from the DBLP computer science bibliography database \cite{BensonASJK18}. This dataset consists of 3,700,681 scientific articles published between the years 2000 and 2016, as well as the list of authors of each article, for a total of 1,930,378 authors. Using this dataset, we build separately a graph, a simplicial complex, and a hypergraph directly from the data for each year contained in the dataset. In each representation, we measure the degree of each node, using the definitions in Section \ref{sec:methods}. Figure~\ref{fig:results1}a contains a scatter plot showing the degree of each node as measured in the different representations corresponding to year 2016. In this dataset, the degree in any representation is positively correlated to the degree in any other representation, though progressively less so as the degree of the node decreases. This result means that the different representations often agree more on which nodes have the largest degrees than on which nodes have small degree. This is important to keep in mind, especially in studies that make claims about the nodes of small degree, which often outnumber those with large degree.

To see how this correlation changes over time, for each year we calculate the Spearman rank correlation coefficient, which quantifies the similarity in node degree rankings between two representations. The Spearman rank correlation coefficient is equal to $1.0$ when the rankings are equal, and $-1.0$ when the rankings are exactly reversed. Shown in Fig.~\ref{fig:results1}b, we measure this coefficient for each year and each pair of representations. We observe the highest correlation between degrees calculated from the simplicial complex representation (participation in maximal simplices) and degrees calculated from the hypergraph representation (number of hyperedges in which a node participates), which is likely due to the fact that these two representations both encode the polyadic relations in the dataset. This result suggests that relatively few papers authored by a subset of the authors of another paper were written. We also observe that the degrees calculated from the graph representations show a comparatively low correlation to the node degrees calculated from the other representations. In particular, the correlation between the graph and the hypergraph drops below $0.5$ in some years, signaling a very different result when ranking nodes by graph degree or by hypergraph degree. Our observations imply that, in this dataset, we should be careful when making broad claims regarding the degree of nodes, especially those with few observed relations (i.e. small degrees), as each representation may yield different results that must be interpreted accordingly.

\paragraph*{} In summary, we have used this example to illustrate each step of the workflow from Figure~\ref{fig:org_methods}, as well as to show that using different representations of the same dataset may yield measurements that are at odds with each other, even in the simple case of measuring node degree.

\subsection{Email communications}

In our next example we again follow the the workflow of Fig.~\ref{fig:org_methods}, but more succinctly. While in the previous example we discussed multiple types of dependencies, variations in data availability, and differing research questions, here we provide an example of a seemingly straightforward analysis on a dataset of emails. 

We start by considering the following scenario. Suppose Ana works at a company and is tasked with improving communication and cohesiveness between teams in the workplace. Ana works at a big company that contains many teams in diverse areas, so she decides to prioritize her involvement by focusing on average team communication via an easily accessible medium such as email. Concretely, Ana wants to evaluate how well each team integrates with all members of the company, which translates to evaluating the average clustering coefficient of each team. For this purpose, Ana has collected all the internal email communications. She decides to operationalize her task as follows. First, if a set of at least $5$ people have all received the same email at the same time, she will assume they must be working together as a team. Second, she decides to focus on emails with at most $25$ participants, as emails with more than 25 participants are likely company-wide communications that do not involve a single team working together. Third, having identified a team, she will quantify the team's cohesiveness by averaging the clustering coefficient of each member in the team. Note that in this system we represent no spatial dependencies as email allows instant communication regardless of geographical location. Further, Ana only cares about the teams and communication that have already occurred, and is not hoping to predict communication in the future, so for the presented analysis on aggregate communication she does not need to incorporate any temporal dependency that might exist within the system.


To follow up with her plan, Ana needs to choose a formalism with which to encode the data, as well as how to measure the clustering coefficient. If she chooses to encode the system as a graph where each employee is a node and each edge joins two nodes if they simultaneously received the same email, then she may use the clustering coefficient defined in Section \ref{sec:methods-for-graphs}. Alternatively, she may choose to build a hypergraph where each node is an employee and each hyperedge denotes a single email, and use the clustering coefficient defined in Section \ref{sec:methods-for-hypergraphs}. A priori, one might expect that measuring the average clustering of a set of nodes in the graph is highly correlated to measuring the same quantity in the hypergraph. However, we will see that the subtle differences in definitions lead to varying results.

\begin{figure}
    \centering
    \includegraphics[width=0.8\textwidth]{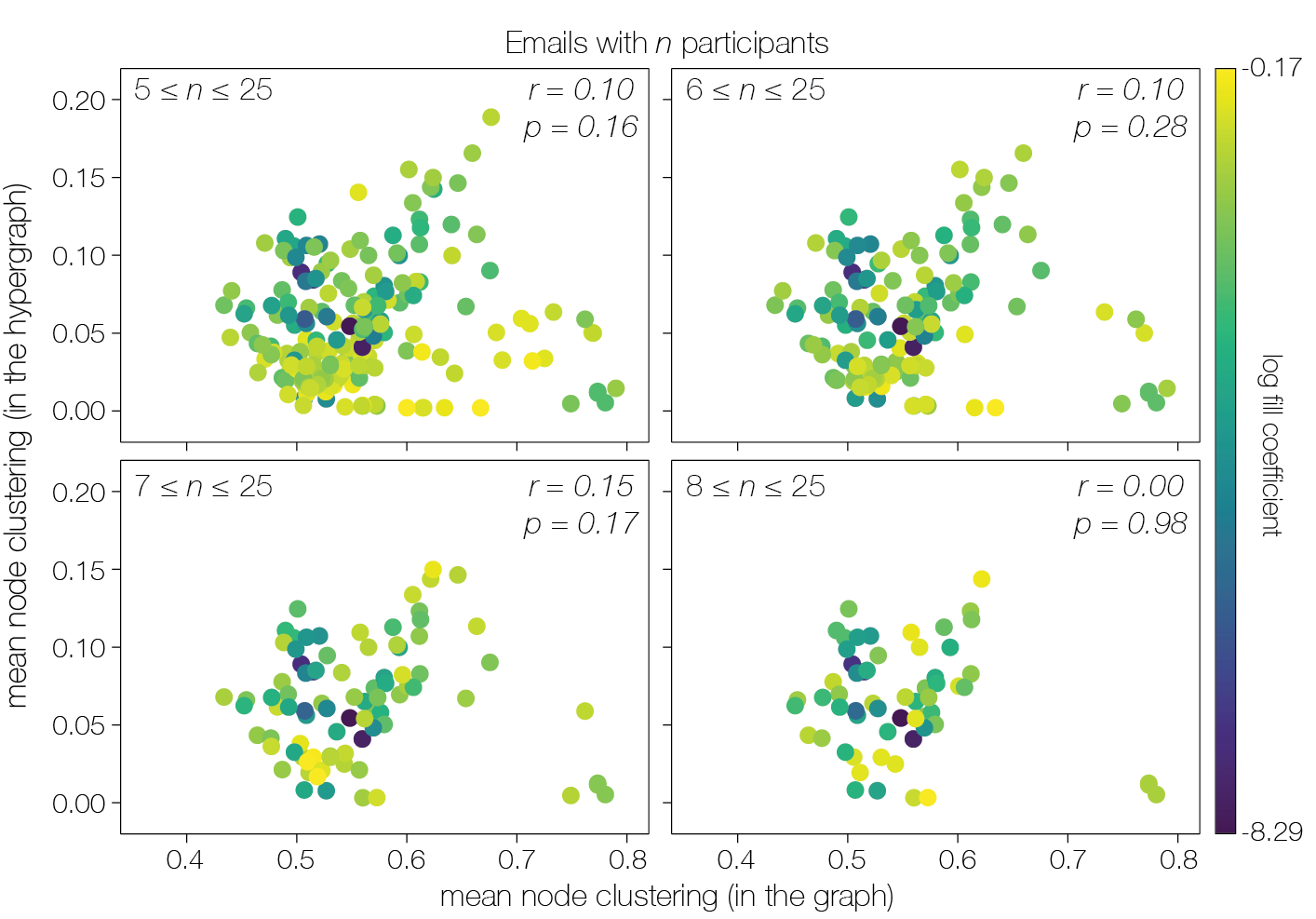}
    \caption{Graph clustering and hypergraph clustering coefficients are loosely related. For different ranges of hyperedge cardinality, the average clustering coefficient of nodes within a hyperedge calculated with the projected graph definition is compared to the average clustering coefficient calculated from the hypergraph representation. Scatterplot points are colored by the $\log(\text{fill coefficient)}$.}
    \label{fig:enron}
\end{figure}

For this example, we use a dataset of email communications \cite{BensonASJK18}, containing $10,883$ emails among $148$ employees of a company, from March 1999 to October 2002. Each email has a corresponding set of participants which includes the sender and all recipients. In Figure~\ref{fig:enron} we see the results of Ana's analysis on this dataset, using both a graph and a hypergraph representation of these data. Each marker represents an email with between $n$ and $25$ participants, for $n = 5,6,7,8$. Each email is located according to the average clustering coefficient of the email's participants, as measured in the graph (horizontal axis) and in the hypergraph (vertical axis). In the top left panel we can see that there is very little correlation between these two quantities for teams of at least $5$ people (Spearman rank correlation coefficient $r=0.1$, and associated p-value $p=0.16$). These results show that ranking teams of employees using these two different clustering coefficients yields very different results. Consequently, if Ana wants to prioritize teams in order of how much she needs to intervene, in other words by ranking the teams according to average clustering coefficient, then the graph and hypergraph representations will recommend very different courses of action. Indeed, Ana would need to allocate her resources in entirely different ways depending on which representation she chose. This result does not change if Ana chooses to focus on teams of at least $6, 7$, or $8$ people, as shown in Fig~\ref{fig:enron}.

Next, Ana decides to distinguish between team cohesiveness with the company and intra-team cohesiveness. That is, do those teams that communicate well with everyone in the company also have robust within-team communication? A team that has excellent internal communication would have a high fill coefficient, which recall measures the fraction of possible smaller hyperedges that exist between the nodes of a hyperedge (see Section \ref{sec:methods-for-hypergraphs}). In order to answer this question Ana calculates the fill coefficient of each team and adds this information as color on her scatterplots (Fig.~\ref{fig:enron}). By eye her results show no relationship between a team's cohesiveness with the company (clustering coefficient calculated from the graph or hypergraph representation) and a team's internal cohesiveness (fill coefficient).

Together, these experiments illustrate that even in the case when a) the research question is fixed, b) the researcher has access to the full dataset, and c) there are little-to-no interactions among different types of dependencies, the choice of representation alone may still yield different insights by virtue of the different assumptions made by each (here dyadic \textit{versus} polyadic interactions).


\section{Applications}\label{sec:applications}

We can naturally encode myriad systems in the real world with at least one of the formalisms discussed in this work. Still, often we focus on analyzing a system from a \emph{particular} perspective and spend less time imagining how alternative analysis pipelines may be more revealing -- or indeed more true to the system -- than the currently used pipeline. Here we consider the alternative perspectives offered by the dependencies, formalisms, and challenges we have discussed in this paper.

First we consider the brain. The brain can be naturally conceived of as a system of individual parts that work together to form large functional units at different scales: neurons work together to communicate with each other forming co-firing patterns called \emph{code words} \cite{curto2017makes}, multiple neuronal populations collaborate in order to plan and evaluate trajectories \cite{olafsdottir2018role}, and entire brain regions work in unison to form functional networks \cite{gallen2019brain}. Scientists have successfully studied the brain by encoding it at any one of these scales using the representations discussed in this work \cite{betzel2017multi}. For example, with graph representations researchers found the brain to exhibit small-world \cite{bassett2006small,bassett2017small,muldoon2016small}, modular architecture \cite{sporns2016modular,gallen2019brain}, and hubs \cite{achard2012hubs,achard2006resilient}. Using the simplicial complex representation, at the larger scale cavities in the structural adult brain were observed \cite{petri2014homological,sizemore2018cliques}, and at a smaller scale researchers detected the geometric structure of pyramidal neuron firing patterns \cite{giusti2015clique}. Finally, research employing a hypergraph representation has identified functional hub hyperedges \cite{wang2012naive}, characterized types of hyperedges in developing children \cite{gu2017functional}, and tracked changes in brain organization over both short \cite{bassett2014cross,davison2015brain} and long time scales \cite{davison2016individual}.
 
Looking to the future, one particularly little-understood aspect of the brain is the impact of temporal dependency. How do specific relations affect the existence of any other relations in the future? For example, can a brain transition from any arbitrary state to any other state \cite{cornblath2020temporal}, or is its future activity bound by its past activity \cite{barnes2009endogenous,wegner2017information}? Though the field has used temporal networks to investigate time-varying activity, at the time of writing we did not find the inclusion of temporal dependencies within the representation. We suggest the application of higher-order network representations to deepen our understanding of temporal dependencies in this complex system. Additionally, at all scales the brain is spatially embedded \cite{stiso2018spatial}, and previous work has shown that the strength of connections between brain regions often depends on the euclidean distance between them \cite{horvat2016spatial,rivera2011wiring}. Often the analysis pipeline involves comparing any computed results on the empirical data against a spatially-embedded null model \cite{betzel2017modular}, or co-modeling the spatial relations and other relations such as by examining Rentian scaling \cite{bassett2010efficient,how2018evidence,sadovsky2014mouse}. While these approaches do help to determine dependence of spatial features on other features, we suggest taking an additional step to directly encoding spatial dependencies in the formal representation of the data. For example, multilayer representations or sheaves encoding position information may be of help here.

Next we consider transportation, which is another well-studied system in network science. Transportation networks come in two different types: systems where the movement is done along fixed routes (such as roads, train tracks, power lines, or airline paths \cite{rodrigue2016geography,bast2016route,seaton2004stations, pagani2013power,colizza2006role}), and systems where the movement is done freely through (outer) space \cite{ross2001lunar}. Among these, analyses specifically of public transportation networks have incorporated multilayer networks \cite{von2009public} or variations thereof including internal node structure \cite{shanmukhappa2018multi}, and have also evaluated system-specific measures that include spatial organization \cite{von2007network}. Analyses involving temporal representations have included investigating congestion clusters in road networks \cite{rempe2016spatio} and how to alleviate them \cite{ji2011spatial}. These analyses usually consider spatial and temporal dependencies by assigning weights to the representation's relations associated to distances or travel times \cite{porta2006network,scheurer2008spatial}. Importantly, studies are beginning to encode the temporal dependency in the representation itself, as higher order networks have revealed these temporal dependencies in data from global shipping and web browsing \cite{xu2016representing}.

The subset dependency is studied far less often than other types of dependencies in transportation systems. For example, in a public transport system, if relations are defined among $k$ stations if there exists a route $X$ that stops at all $k$ stations, then certainly any subset of those $k$ stations must also be related by route $X$. Alternatively a subset dependency may or may not exist within traversed paths. For example, perhaps we observe paths of length $k$ but we do not observe smaller sub-paths. How might the identification or inclusion of subset dependencies within the transportation system improve our ability to prevent system failures or predict future activity? Additionally, we suggest further investigation of polyadic relations in these systems with simplicial complexes or hypergraphs where appropriate, as these representations may elucidate previously hidden system properties.



Finally we consider applications in cellular systems composed of any subset of proteins, genes, regulatory units such as enhancers, epigenetic factors, and more \cite{fionda2019networks,alon2007network}. Most commonly the field studies system fragments such as genetic regulatory networks (GRNs) \cite{de2002modeling,walhout2011gene,emmert2014gene} and protein-protein interaction networks (PPINs) \cite{de2010protein,raman2010construction}. Unlike the above two examples, this application differs in that only in rare situations can real-world interactions be observed. Consequently, tremendous effort focuses on network reconstruction from data, i.e. inferring the interactions from indirect measurements \cite{oates2012network,vinci2019graph,albert2007network}. Temporal information such as fluctuations in RNA counts \cite{spies2015dynamics} and spatial information such as co-localization \cite{mardakheh2017proteomics} can be used to reconstruct the network of interactions. Often, for example in the system fragment of proteins and protein complexes, polyadic relations exist and have been encoded using simplicial complexes or hypergraphs \cite{ramadan2004hypergraph}. Multilayer representations have also been used for representing multiple biological layers important in disease \cite{halu2019multiplex} and for inferring protein function \cite{zhao2016efficient}.

The difficulties associated with macromolecule-interaction systems create an enticing problem for developing system representations. Since the existence of interactions can rarely be observed directly, perhaps a representation with weights indicating the probability of the existence of each relation might be a useful alternative. In such a case, one might consider studying an ensemble of graphs, simplicial complexes, or hypergraphs instead of only one, and differentiating among them based on the likelihood of each one being a faithful representation of the real system. Additionally, though polyadic relations are known to play an important role in these systems \cite{taylor2015higher}, simplicial complex and hypergraph representations are more rarely employed (examples include \cite{estrada2018centralities,shnier2019persistent,klamt2009hypergraphs}. Finally, temporal fluctuations are becoming easier to record in these systems \cite{savulescu2019dypfish}, which poses the opportunity to directly study temporal dependencies.

\section{Discussion and Conclusion}\label{sec:conclusion}

\begin{figure}
    \centering
    \includegraphics[width = \textwidth]{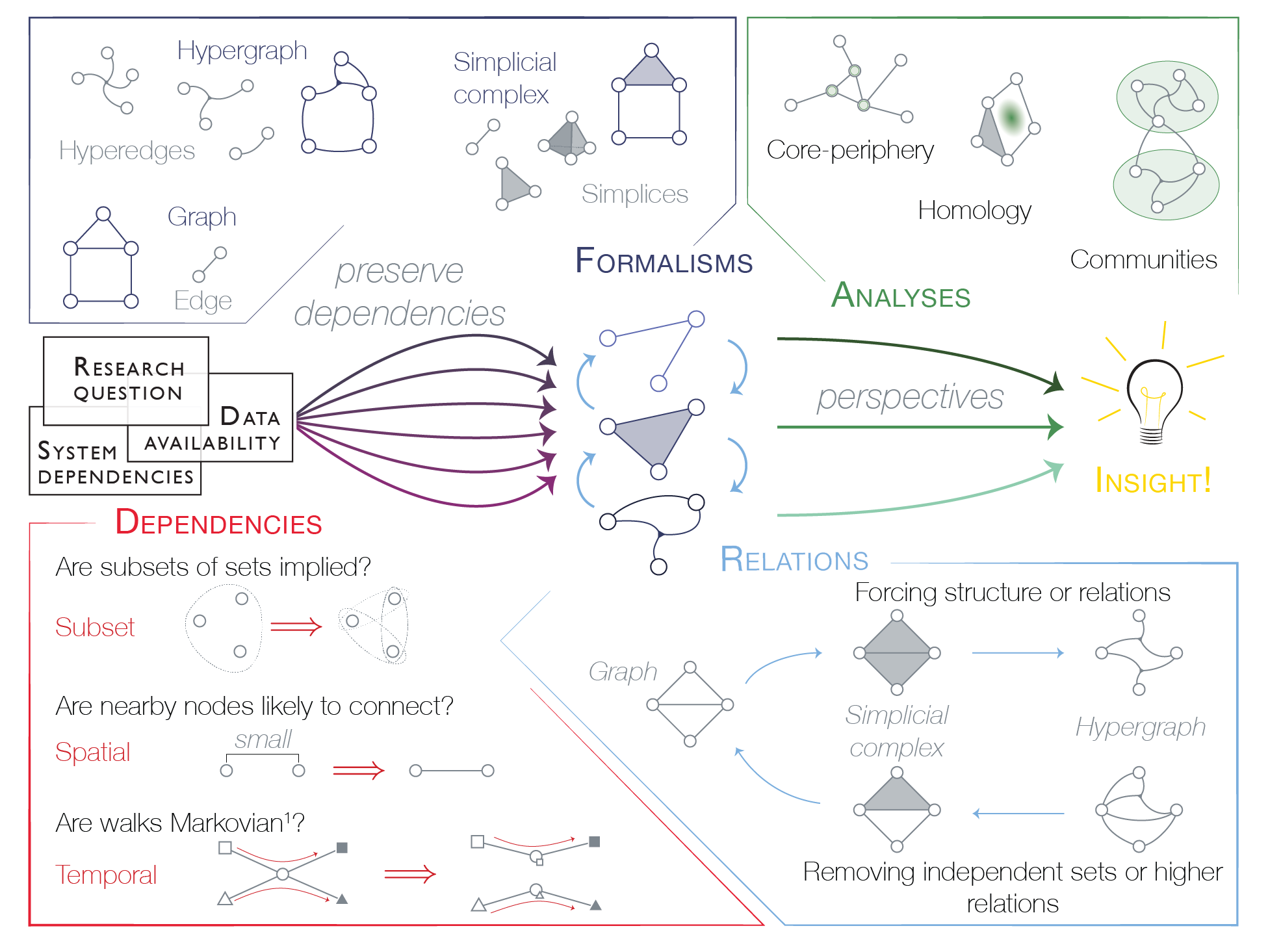}
    \caption{Complex system analysis pipeline discussed in this paper. We suggest beginning with considering how the research question, data availability, and system dependencies may influence downstream analyses. We discussed subset, spatial, and  temporal dependencies (red). We suggest choosing a formalism (navy) that preserves and respects dependencies within the system and data. Formalisms themselves are mathematically related (light blue), but switching formalisms after the initial data encoding can result in making inaccurate assumptions or forgetting independent relations. Finally, choosing the appropriate analysis method for the system and representation (green) after all other steps have been performed carefully can offer insight into the system's behavior, structure, or function.}
    \label{fig:flow_conclusion}
\end{figure}

In this work we examined each step of a data analysis pipeline suitable for studying complex systems (Fig.~\ref{fig:flow_conclusion}). We first discussed system dependencies which can manifest in different flavors including but not limited to temporal, subset, and spatial. We then defined common complex system formalisms and their underlying assumptions, as well as which dependencies they encode. We discussed the mathematical relationships between formalisms, and how information can be lost (or imputed) as we convert data from one formalism to another. Finally we offered analysis examples in order to underscore the importance of dependencies, careful choice of representation, and analysis techniques in studying complex systems.

The main message of our work is that there is no perfect way to analyze a system, and that studying two different systems may require two entirely different pipelines. That is, the modeling decisions made while studying one dataset compiled from a system will not necessarily carry over to another system or, indeed, not even to another dataset extracted from the same system. In contrast, we see many studies apply certain pipelines for seemingly no other reason than because they are common within a certain field. Instead, we recommend that each new system and dataset be individually evaluated and investigated, and each assumption and pipeline decision be made in accordance with the concepts discussed here. 
More specifically, and following Fig~\ref{fig:flow_conclusion}, we suggest designing pipelines based on the system and system dependencies, any external dependencies that may be induced by the data type or data collection method, and the limits of the system fragment under study. From there, we suggest choosing a formalism that best fits the data, the research question, and the system itself, even if it requires using a new formalism or extension that is outside of what is customary. Finally, we recommend choosing carefully the specific methods, measurements, and analyses done on the chosen representation, and keeping in mind that their results may be biased by the choices made in the previous stages. Different choices at each of these steps may ultimately yield results that are at odds with the results yielded by other choices. Only after respecting the system's dependencies and unique qualities through proper representation and analysis methods will we uncover novel insight into the system under study.


Though here we present only a first attempt to unify the application of complex systems analyses, we hope that the drive for more accurate representations will continue to push the field both forward \emph{and} closer together through multiplying collaborations. We imagine that complex systems researchers in the future may each have a slew of representations along with carefully chosen computations that respect the dependencies one finds within the system. By continuing this discussion, the separate areas of science that use complex systems analyses will together identify what is missing from current formalisms, create more insightful analyses, and generate novel techniques.

\section{Acknowledgments}

First and foremost, we wish to acknowledge the colleagues, coauthors, mentors, and mentees who have shaped our perspective on this subject. 
Finally, we acknowledge critical financial support that allowed us to devote time to this work. ASB and DSB acknowledge support from the Army Research Office (Falk-W911NF-18-1-0244, Grafton-W911NF-16-1-0474, DCIST- W911NF-17-2-0181), the National Science foundation (PHY-1554488, IIS-1926757), and the Paul G. Allen Family Foundation. LT and TER were supported in part by the National Science Foundation (IIS-1741197) and by the Combat Capabilities Development Command Army Research Laboratory (under Cooperative Agreement Number W911NF-13-2-0045). The views and conclusions contained in this document are those of the authors and should not be interpreted as representing the official policies, either expressed or implied, of the Combat Capabilities Development Command Army Research Laboratory or the U.S.~Government. The U.S.~Government is authorized to reproduce and distribute reprints for Government purposes not withstanding any copyright notation here on

\section{Citation diversity statement}
Recent work in several fields of science has identified a bias in citation practices such that papers from women and other minorities are under-cited relative to the number of such papers in the field \cite{dworkin2020extent,maliniak2013gender,caplar2017quantitative,chakravartty2018communicationsowhite,thiem2018just,dion2018gendered}. Here we sought to proactively consider choosing references that reflect the diversity of the field in thought, form of contribution, gender, and other factors. Gender bias can arise due to explicit and implicit bias against a person’s known gender as a woman, or due to explicit or implicit bias against a person carrying a name commonly used by women \cite{macnell2015s,paludi1985s,moss2012science}. To evaluate the former (bias according to known gender), we obtained predicted gender of the first and last author of each reference using pronouns affiliated with them online  or pronouns known by personal friendships; by this measure (and excluding self-citations to the first and last authors of our current paper), our references contain 45\% man(first)/man(last), 12\% man/woman, 11\% woman/man, 13\% woman/woman, 0\% non-binary , and 19\% unknown categorization. This method is limited in that pronouns may not be indicative of gender identity, and may not be consistent across time or environment. To evaluate the latter (bias according to a gendered name), we used databases that store the probability of a name being carried by a woman; by this measure (again excluding self-citations), our references contains 60\% man/man names, 12\% man/woman names, 11\% woman/man names, 10\% woman/woman names, and 7\% unknown categorization \cite{zhou2020gender,dworkin2020extent}. This method is limited in that it cannot account for intersex, non-binary, or transgender people. We look forward to future work that could help us to better understand how to support equitable practices in science.

\newpage
\bibliographystyle{plain}
\bibliography{references}

\end{document}